\documentclass[10pt,a4paper]{article}
\usepackage[english]{babel}
\usepackage[latin1]{inputenc}
\usepackage{amsfonts,amsbsy,bm,euscript,mathrsfs}
\usepackage{amssymb,stmaryrd,faktor}
\usepackage[tbtags]{amsmath}
\usepackage{bbm}
\usepackage{graphicx}
\usepackage[title,titletoc]{appendix}
\usepackage[bookmarks=true,colorlinks=true,linkcolor=blue,citecolor=blue,urlcolor=blue,bookmarksnumbered]{hyperref}

\usepackage[dvipsnames]{xcolor}
\usepackage{soul}

\usepackage{dsfont}
\usepackage{collref}
\usepackage{lmodern}
\usepackage{mathrsfs}
\usepackage{mathtools}
\usepackage{bbm}
\usepackage{braket}
\usepackage{slashed}
\usepackage{float}
\usepackage{graphicx}
\usepackage{booktabs}
\usepackage{subfig}
\usepackage{tikz}
\usepackage[symbol]{footmisc}
\usepackage{hyperref}
\usepackage{cleveref}
\usetikzlibrary{plotmarks,calc,decorations,decorations.pathmorphing}

\makeatletter
\def\tagform@#1{\maketag@@@{\normalsize(#1)\@@italiccorr}}
\makeatother

\numberwithin{equation}{section}

\makeatletter
\renewcommand\section{\@startsection {section}{1}{\z@}
{-3.5ex \@plus -1ex \@minus -.2ex}
{2.3ex \@plus.2ex}
{\normalfont\Large\bfseries}}
\renewcommand\subsection{\@startsection{subsection}{2}{\z@}
{-3.25ex\@plus -1ex \@minus -.2ex}
{1.5ex \@plus.2ex}
{\normalfont\large\bfseries}}
\makeatother

\newcommand{\arXivlink}[1]{\href{http://arXiv.org/abs/#1}{arXiv:#1}}

\newcommand{\alg}[1]{\mathfrak{#1}}

\usepackage{blkarray}

\begin{document}

\thispagestyle{empty}
\begin{flushright}\footnotesize\ttfamily
DMUS-MP-24-07
\end{flushright}
\vspace{2em}

\begin{center}

{\Large\bf \vspace{0.2cm}
{\color{black} \large Boundary Bethe ansatz in massless $AdS_3$}} 
\vspace{1.5cm}

\textrm{Daniele Bielli$^{a}$, Vasileios Moustakis$^{b}$ and Alessandro Torrielli$^{b}$ \footnote[1]{\textit{E-mail:} \texttt{d.bielli4@gmail.com, \{v.moustakis@,  a.torrielli@\}surrey.ac.uk}}}
\vspace{0.8cm}
\\
\vspace{0.3cm}
$^a$ \textit{High Energy Physics Research Unit, Faculty of Science,
Chulalongkorn University,
\\
Bangkok 10330, Thailand}
\\
$^b$ \textit{Department of Mathematics, University of Surrey, Guildford, GU2 7XH, UK}
\vspace{0.3cm}




\end{center}

\vspace{2em}

\begin{abstract}\noindent 
 
We construct the boundary algebraic Bethe ansatz for the $AdS_3 \times S^3 \times T^4$ integrable reflection problem restricted to the massless sector. We derive the double-row monodromy and find the appropriate formulation of the {{} dual} equation of Sklyanin's. We perform the algebraic Bethe ansatz {{} and} obtain the RTT (more properly RTRT) relations, from which the spectrum is obtained by repeated action of the $B$ operator on the pseudovacuum. We obtain the Bethe equations by cancelling the unwanted terms, and {{} prove} the exact formulae for any number of quantum particles and magnonic excitations.

\end{abstract}

\newpage

\overfullrule=0pt
\parskip=2pt
\parindent=12pt
\headheight=0.0in \headsep=0.0in \topmargin=0.0in \oddsidemargin=0in

\vspace{-3cm}
\thispagestyle{empty}
\vspace{-1cm}

\tableofcontents

\setcounter{footnote}{0}

\section{Introduction}

In this paper we shall probe the integrable structure of the massless sector of string theory on $AdS_3 \times S^3 \times T^4$ \cite{Bogdan} (for reviews, the reader can refer to \cite{rev3,Borsato:2016hud}). Our goal is to enhance our understanding of this fascinating integrable system, which is itself enriching the armamentary of techniques and effects which AdS/CFT features \cite{Beisertreview}. The history of this sub-genre is by now long \cite{OhlssonSax:2011ms,seealso3,Borsato:2012ss,Borsato:2013qpa,Borsato:2014hja,Borsato:2013hoa} and its most important new ingredient, with respect to the higher-dimensional $AdS$ cousins, is in our opinion the very presence of the massless excitations \cite{Sax:2012jv}. A proper interpretation of these excitations is usually gained by calling in the game the notion of massless $S$-matrices \cite{Zamol2,Fendley:1993jh,DiegoBogdanAle}, see also \cite{Lloyd:2013wza} and \cite{Ben}. The massive sector has been submitted to the method of  Quantum Spectral Curve (QSC) in \cite{QSC}, see also \cite{Cavaglia:2022xld}, but a full understanding of the massless sector within the QSC is still elusive. The papers \cite{AleSSergey}, see also \cite{Seibold:2022mgg}, have revisited the problem of $AdS_3 \times S^3 \times T^4$, with a new formulation of the dressing factors and a derivation of the mirror TBA equations. More work can be found in \cite{recent}.

In the paper \cite{Bielli:2024xuv}, we developed the boundary scattering \cite{GhoshalZamo} for the massless sector of $AdS_3 \times S^3 \times T^4$, following onto the steps of the massive version  \cite{Prinsloo:2015apa}, but encountering a host of new phenomena peculiar to the massless situation. We refer to \cite{Prinsloo:2015apa} for references on boundary reflection matrices, primarily constructed for $AdS_5$. The greatest technical advantage of the massless sector is the existence of a new set of variables \cite{gamma1,gamma2} which allows to recast the scattering problem in relativistic terms \cite{DiegoBogdanAle,AleSSergey}. The use of the relativistic variables allowed us to find general solutions to the boundary Yang-Baxter equation for the singlet boundaries - these are boundaries without degrees of freedom, which cannot change the polarisation (boson/fermion) of the particle which they reflect. We also understood the importance of the boundary co-ideal (sub)algebra condition in ``correcting" the choice of co-product by attaching suitable rescaling factors to the algebra generators, which allowed us to find the various vector boundaries solutions - these are boundaries which {{} can} change the polarisation of a particle, and effectively behave like a magnon stuck on the boundary with zero group velocity. 

What we perform here is the boundary Bethe ansatz, the natural next step to the development of the complete theory of boundary scattering which we have previously constructed. {{} The application of the algebraic Bethe ansatz to supersymmetric integrable systems with a boundary is the object of a vast literature\footnote{{} We thank the anonymous referee for extremely useful suggestions.}. The reader can consult for example \cite{Bedu,Bracken:1997zww,AppelVlaar,Angela,Foerster:1993fp,Slav,Ge,Gonzalez-Ruiz:1994loj,Grab,Links,Gould,Martins:1999jbx,Shiro,Yao,Zhou,Zhou:1998joy}.}  We focus on the singlet boundary and the $L$ representation, as defined in \cite{Borsato:2014hja}. The {{} level of technicality} is sufficient {{} in our opinion} to make this work self-contained, in particular as we {{} pedagogically} setup all the stages of the  algebraic Bethe ansatz programme \cite{Sklyanin:1988yz}. 
The paper constructs the double-row monodromy matrix, then the commuting transfer matrix and the traditional operators $A,B,C,D$, which build the magnons by applying the $B$ operators on the pseudovacuum, finally using the RTT (or, {{} more properly, RTRT}) relations to prove the eigenstate's property. We recall that the pseudovacuum is not necessarily the true {{} vacuum} 
but it is merely a suitable highest/lowest-weight state from which to generate the entire spectrum by repeated creation {\it via} $B$'s. Cancelling the unwanted terms, as always, will give us the auxiliary Bethe equations. The main obstacle to overcome in this context is the derivation of the so-called {\it left} or {\it dual} equation of Sklyanin's - {{} the construction of this equation is the subject of a vast literature (see for example \cite{Bracken:1997zww}), and there exist supersymmetric versions of it. We have independently obtained an appropriate dual equation tailored to the case at hand, which allowed to promptly proceed with the formalism. The} general expressions for any number of physical sites and any number of magnons have been {{}first constructed by extrapolating from small number of particles/sites, and then proven, thanks to a number of identities which the various massless $S$ matrices of $AdS_3$ satisfy and relying on the familiar armamentary of algebraic Bethe ansatz proofs\footnote{{} We are indebted to the anonymous referee for their encouragement and advice.}.}

{{}
\section{Notation and conventions}\label{sec:notation-and-conventions}

In this section we summarise our notation and conventions for graded tensor products, $R$-matrices and states. There are several excellent expositions of supersymmetric conventions for Hopf algebras (see for instance \cite{Yao}) - we prefer to make completely explicit our own conventions for completeness and for ease of reading.  

We start by defining the vector space $V = \mathbbmss{C}^{(1|1)}$, carrying the basic  representations which we are concerned with. The states are denoted by
\begin{eqnarray}
    |\phi\rangle = |1\rangle \quad \text{boson (with grading } |1|=0)
    \quad \quad , \quad \quad |\psi\rangle = |2\rangle \quad \text{fermion (with grading } |2|=1) \,\, .
\end{eqnarray}
These states form the fundamental (defining) representation of the Lie superalgebra $\mathfrak{psu}(1|1)$, which is the basic building block of the total symmetry of the model. We also use the matrices 
\begin{eqnarray}
E_{ij} \,\, , \qquad \text{for} \qquad  i,j \in \{1,2\} \,\, ,  
\end{eqnarray}
with all zeroes except 1 in row i, column j, which act on states as
\begin{eqnarray}
    E_{ij} |k\rangle = \delta_{jk} |i\rangle \,\, , \qquad \text{for} \qquad i,j,k \in \{1,2\} \,\, ,
\end{eqnarray}
and have grading $|E_{ij}|=(|i|+|j|) \, \mbox{mod} \, 2$. The tensor product is graded as well, therefore
\begin{equation}
\begin{aligned}
& (E_{ij} \otimes E_{kl})  (E_{ab} \otimes E_{cd}) = (-1)^{(|a|+|b|)(|k|+|l|)} \delta_{ja} \delta_{lc} E_{ib} \otimes E_{kd} \,\, , 
\\
&(E_{ij} \otimes E_{kl}) |m\rangle \otimes |n\rangle = (-1)^{|m|(|k|+|l|)} \delta_{jm} \delta_{ln} |i\rangle \otimes |k\rangle \,\, .
\end{aligned}
\end{equation}
In writing $R$-matrices we always order our basis as $\{ |11\rangle,|12\rangle,|21\rangle,|22\rangle \}$, where we used the simplified notation $|ij\rangle \equiv |i\rangle \otimes |j\rangle$. Our convention for the "opposite" operation ${}^{op}$ is $R^{op} = \Pi (R)$, where $\Pi$ is the graded permutation acting on tensor products of matrices, namely \begin{eqnarray}
\Pi\Big( E_{ab}\otimes E_{cd}\Big) = (-1)^{(|a|+|b|)(|c|+|d|)} E_{cd} \otimes E_{ab} \,\, .   \label{usa} 
\end{eqnarray}
The same result can be obtained by using the permutation operator $P_s$ on states, defined by
\begin{eqnarray}\label{Ps-definition}
P_s =  (-1)^{|j|} E_{ij} \otimes E_{ji} \,\, ,\qquad \text{and acting as} \qquad P_s (|a\rangle \otimes |b\rangle) = (-1)^{|a||b|} |b\rangle \otimes |a\rangle  \,\, .
\end{eqnarray}
Unless otherwise specified, summation over repeated indices is understood and a generic $R$-matrix reads
\begin{equation}\label{generic-R-matrix-expansion}
R = r^{abcd} E_{ab} \otimes E_{cd} \,\, .
\end{equation}
It is indeed easy to show that
\begin{eqnarray}
 \Pi ( R ) = P_s R P_s.   
\end{eqnarray}
The proof goes as follows: from the action of $\Pi$ in \eqref{usa} we have
\begin{eqnarray}\label{Pi-on-R}
    \Pi ( R ) = r^{abcd} \, (-1)^{(|a|+|b|)(|c|+ |d|)} E_{cd} \otimes E_{ab} \,\, ,
\end{eqnarray}
while the definition \eqref{Ps-definition} of $P_s$ leads to
\begin{equation}
\begin{aligned}
P_s R P_s &= (-1)^{|j|+|n|} r^{abcd} \, [E_{ij} \otimes E_{ji}][E_{ab} \otimes E_{cd}][E_{mn}\otimes E_{nm}]
\\
&=r^{abcd} \, (-1)^{|j|+|n| +(|a|+|b|)(|i|+|j|) + (|j|+|d|)(|m|+|n|)} \, \delta_{ja} \, \delta_{ic} \, \delta_{bm} \, \delta_{dn} \, E_{in} \otimes E_{jm}  
\\
&=r^{abcd} \, (-1)^{|a|+|d| +(|a|+|b|)(|c|+|a|) + (|a|+|d|)(|b|+|d|)} \, E_{cd} \otimes E_{ab}
\\
&=r^{abcd} \, (-1)^{(|a|+|b|)(|c|+|d|)} \, E_{cd} \otimes E_{ab},
\end{aligned}
\end{equation}
where in the last line we have used that $|a||a|=|a|$ and $(-1)^{2|a|} = 1$ for any $a \in \{0,1\}$.

As from \eqref{generic-R-matrix-expansion}, the $R$-matrix acts on two copies of the basic vector space
\begin{eqnarray}
R : \quad V_1 \otimes V_2 \longrightarrow V_1 \otimes V_2 \,\, , 
\end{eqnarray}
where $V_i$ are identical copies of $\mathbbmss{C}^{(1|1)}$, carrying representations which are characterised by distinct parameters $\gamma_i$, introduced below. In equations involving multiple tensor products, we indicate explicitly the spaces on which the $R$-matrix acts non-trivially. For instance, we denote by $[R]_{1,2}$ an $R$-matrix acting on spaces $V_1$ and $V_2$. Our standard ordering for boundary transfer matrices is $0-1-2-...-(N-1)-N-(N-1)-...-2-1$ (abbreviated by $0-1-2-...-N$), with $0$ denoting an auxiliary space and $1-...-N$ denoting physical spaces. In cases where two auxiliary spaces need to be taken into account, we denote them by $0$ and $0'$, with the standard ordering modified as $0-0'-\mbox{[physical spaces]}$. As discussed below, opposite $R$-matrices often appear in the definition of transfer matrices and we now wish to make our notation fully explicit. For example, in the case of $[R^{op}_{+-}]_{0,2}$, with the chirality subscripts $+$ and $-$ to be introduced below and respectively associated to the spaces $0$ and $2$, the $R$-matrix $R_{+-}(\gamma)$ is first decomposed on the basis as in \eqref{generic-R-matrix-expansion} $R_{+-}(\gamma) = r_{+-}^{abcd}(\gamma) \, E_{ab} \otimes E_{cd}$ and then acted upon by $\Pi$ as in \eqref{Pi-on-R}. Finally, the action on any space different from $0$ and $2$ is made trivial by taking tensor products of identity matrices. The final result takes the form
\begin{eqnarray}
[R^{op}_{+-}]_{0,2} \equiv r_{+-}^{abcd}(\gamma) \, (-)^{(|a|+|b|)(|c|+|d|)} \, E_{cd} \otimes \mathbbmss{1} \otimes E_{ab} \otimes \mathbbmss{1} \otimes ... \otimes \mathbbmss{1} \,\, ,
\end{eqnarray}
acting on $N+1$ copies of $V$ and non-trivially only on the auxilary space $0$ and the physical space $2$.

Here we should note that even though our definitions will respect the assignment of spaces and their chiralities, later on, particularly in proofs, we will be using accidental identites amongst our $R$-matrices which only hold due to the fact that both chirality spaces are isomorphic to $V$. This also means that, for the purposes of this article where all representation are $2$-dimensional, we can practically  multiply matrices in different chirality-spaces interchangeably. 

In the $AdS_3$ model which we focus on here all particles are massless, and come in two types:
\begin{eqnarray}
&&\mbox{right movers} \qquad E_i = 2h \sin \frac{p_i}{2}>0, \qquad p_i \in (0,\pi), \qquad x_i^\pm = e^{\pm \frac{i p_i}{2}},\nonumber\\
&&\mbox{left movers} \qquad E_i = -2h \sin \frac{p_i}{2}>0, \qquad p_i \in (-\pi,0), \qquad x_i^\pm = -e^{\pm \frac{i p_i}{2}},\label{addi}
\end{eqnarray}
with $i=1,2$ denoting the scattering particles.
In these formulae $h>0$ is the coupling constant of the theory.
The right and left nature of a massless particle is referred to as {\it chirality}, and one sets
\begin{eqnarray}
\mbox{right movers} \leftrightarrow \mbox {chirality +} \,\, , \qquad \mbox{left movers} \leftrightarrow \mbox {chirality -} \,\, .
\end{eqnarray}
The chirality genuinely distinguishes particles, so we denote the $R$-matrices with two signs attached to denote the types of particles which participate to the scattering - and, it being an $R$-matrix, the ordering is always fixed before and after the scattering. 

The scattering matrices are of difference form if one uses the pseudo-relativistic variable, which was introduced in \cite{gamma1,gamma2} and further extended in \cite{Majumder:2021zkr,AleSSergey}: 
\begin{eqnarray}
&&\gamma_i = \log \tan \frac{p_i}{4}<0, \quad\,\,\,\, p_i \in (0,\pi),\nonumber\\
&&\gamma_i = \log \cot \frac{-p_i}{4}>0, \quad p_i \in (-\pi,0).\label{acco}
\end{eqnarray}

We sometimes write actual matrices to denote our $R$-matrices, but this writing is inherently ambiguous. We therefore report here what we mean on the basis (our tensor products being always graded):
\begin{equation}\label{R-matrices}
\begin{aligned}
& R_{++}(\gamma) \!=\! E_{11}\!\otimes\! E_{11} \!-\! \tanh \frac{\gamma}{2} (E_{11}\!\otimes\! E_{22}\!-\!E_{22}\!\otimes\! E_{11}) \!-\! E_{22}\!\otimes\! E_{22} \!+\! \mbox{sech} \frac{\gamma}{2} (E_{21}\!\otimes\! E_{12}\!-\!E_{12}\!\otimes\! E_{21}),
\\ 
& R_{--}(\gamma) \!=\! E_{11}\!\otimes\! E_{11} \!+\! \tanh \frac{\gamma}{2} (E_{11}\!\otimes\! E_{22}\!-\!E_{22}\!\otimes\! E_{11}) \!-\! E_{22}\!\otimes\! E_{22} \!+\! \mbox{sech} \frac{\gamma}{2} (E_{21}\!\otimes\! E_{12}\!-\!E_{12}\!\otimes\! E_{21}),
\\
& R_{+-}(\gamma) \!=\! E_{11}\!\otimes\! E_{11} \!-\! \tanh \frac{\gamma}{2} (E_{11}\!\otimes\! E_{22}\!+\!E_{22}\!\otimes\! E_{11}) \!+\! E_{22}\!\otimes\! E_{22} \!-\! i \mbox{sech} \frac{\gamma}{2} (E_{21}\!\otimes\! E_{12}\!-\!E_{12}\!\otimes\! E_{21}), \\
& R_{-+}(\gamma) \!=\! E_{11}\!\otimes\! E_{11} \!+\! \tanh \frac{\gamma}{2} (E_{11}\!\otimes\! E_{22}\!+\!E_{22}\!\otimes\! E_{11}) \!+\! E_{22}\!\otimes\! E_{22} \!+\! i \mbox{sech} \frac{\gamma}{2} (E_{21}\!\otimes\! E_{12}\!-\!E_{12}\!\otimes\! E_{21}). 
\end{aligned}
\end{equation}
All the relations between $R$-matrices which we will need can be obtained by applying the rules outlined in this section to the $R$-matrices written on the basis. As an example, using (\ref{usa}) we can see that
\begin{eqnarray}
R_{++}(\gamma) = R^{op}_{++}(-\gamma) = R_{--}(-\gamma).    
\end{eqnarray}
Using $\alpha,\beta,\rho = \pm $ the complete set of Yang-Baxter equations for all possible chiralities can be written as
\begin{equation}
[R_{\alpha\beta}]_{12}(\gamma_1 \!-\! \gamma_2) \,  [R_{\alpha\rho}]_{13}(\gamma_1 \!-\! \gamma_2) \,[R_{\beta\rho}]_{23}(\gamma_1 \!-\! \gamma_2) \!=\! [R_{\beta\rho}]_{23}(\gamma_1 \!-\! \gamma_2) \,[R_{\alpha\rho}]_{13}(\gamma_1 \!-\! \gamma_2) \,[R_{\alpha\beta}]_{12}(\gamma_1 \!-\! \gamma_2)  \,\, .
\end{equation}
}
We shall adopt throughout the entire paper the conventions {} of \color{black} \cite{AleSSergey} and \cite{Bielli:2024xuv}.
As anticipated, we will focus exclusively on particles in the $L$ representation. With no risk of confusion we shall hence simplify the notation of \cite{Bielli:2024xuv} suppressing the $L$-indices from $R$- and $K$-matrices. {{} Were we to write matrices instead of components, we would write for instance
\begin{eqnarray}
\label{R++}
&&R_{++}(\gamma) = \begin{pmatrix}1&0&0&0\\0&-\tanh\frac{\gamma}{2}&\mbox{sech} \frac{\gamma}{2}&0\\0&\mbox{sech}\frac{\gamma}{2}&\tanh\frac{\gamma}{2}&0\\0&0&0&-1\end{pmatrix},\qquad \gamma = \gamma_1 - \gamma_2, \qquad \gamma_1,\gamma_2<0.
\end{eqnarray}
We will {{} ignore the so-called  dressing} factors in the first instance - they can be reinstated at the very end {{} as they will just contribute} a multiplying factor to the transfer-matrix {{}eigenvalues}. We also denote by ${}^{st_i}$ the super-transposition in the $i$-th space of the tensor product, with the convention $[E_{ij}]^{st} = (-)^{|i|(|j|+1)}E_{ji}$.} 



\section{The AdS boundary Bethe ansatz}

\subsection{Double-row monodromy and transfer matrix}

We start by recalling the singlet solution to the right-wall boundary Yang-Baxter equation \cite{Bielli:2024xuv}:
\begin{eqnarray}
\label{BYBE_original_form}
&&K_2(\gamma_2) R^{op}_{+-}\Big(\gamma_2 - (-\gamma_1)\Big)K_1(\gamma_1)R_{++}\Big( \gamma_1-\gamma_2 \Big) = \nonumber\\
&&\qquad \qquad R^{op}_{--}\Big((-\gamma_2)-(-\gamma_1)\Big)K_1(\gamma_1)R_{+-}\Big(\gamma_1 - (-\gamma_2)\Big)K_2(\gamma_2),
\label{BYBE}
\end{eqnarray}
where 
\begin{eqnarray}\label{refa}
K(\gamma_i) = \begin{pmatrix}1&0\\0&f(\gamma_i)\end{pmatrix}, \qquad f(\gamma_i) = -i \tanh \Big( \mbox{arccoth} (e^{\gamma_i}) + i c^{L}\Big),
\end{eqnarray}
with $c^{L}$ ($L$ to highlight the $L$ representation) any complex constant and $K_{1},K_{2}$ defined in the usual fashion $K_{1}\equiv K \otimes \mathbbmss{1}$, $K_{2}\equiv  \mathbbmss{1} \otimes K$. We work with regular singlet solutions for the reflection matrix, and normalise the boson entry to $1$.  We can also prove the following re-parameterisation (the second one inspired by \cite{Nepomechie:2008ab}):
\begin{eqnarray}
   f\big(\gamma(p)\big) = -i \tanh \Big[\frac{1}{2} \mbox{arctanh} \Big(\sin \frac{p}{2}\Big) - i c\Big] =  -e^{i \frac{p}{2}} \, \frac{a+x^-(p)}{a-x^+(p)},
\end{eqnarray}
provided that
\begin{eqnarray}
c = \mbox{arctan} (\mbox{cot} \, c^{L}) + k \pi, \qquad k \in \mathbbmss{Z}, \qquad a= 1 + \frac{2}{-1 + \tan c}, \qquad x^\pm(p) = e^{\pm i \frac{p}{2}},    
\end{eqnarray}
in addition to the right-mover formula $\gamma(p) = \log \tan \frac{p}{4}$, $p\in (0,\pi)$. One can also rewrite the above reflection matrix $K$ in such a way that it is manifestly meromorphic in the rapidity $\gamma_i$:
\begin{eqnarray}
 f(\gamma_i) = -i \tanh \Big( \mbox{arccoth} (e^{\gamma_i}) + i c^{L}\Big)  = -i \,\, \frac{+1 + e^{2ic^L}+(e^{2i c^L}-1)e^{\gamma_i}}{-1+e^{2 i c^L}+(e^{2i c^L}+1)e^{\gamma_i}}, \qquad \gamma_i <0.
\end{eqnarray}


\noindent
{{} Following} the general prescription \cite{Sklyanin:1988yz}, we first construct the $T_-$ monodromy matrix. Like the $K$ matrix above, $T_-$ is a solution of the BYBE \eqref{BYBE_original_form} - or \eqref{BYBEa} - 
but it also acts on a ``quantum" space consisting of $N$ fundamental representations tensored together, such that the total space on which $T_-$ acts is $V_0 \otimes V_1 \otimes V_2 \otimes ... \otimes V_{N-1} \otimes V_N$, each $V$ being isomorphic to the fundamental $\{|\phi\rangle,|\psi\rangle\}$ space. We have    
\begin{equation}
\label{Tminus}
\begin{aligned}
&&T_- = [R_{++}]_{0,N}(\gamma_0-\gamma_N)\,[R_{++}]_{0,N-1}(\gamma_0-\gamma_{N-1})...[R_{++}]_{0,2}(\gamma_0-\gamma_2)\,[R_{++}]_{0,1}(\gamma_0-\gamma_1)K_{0}(\gamma_0) 
\\
&&\qquad \qquad \times[R^{op}_{+-}]_{0,1}(\gamma_1+\gamma_0)\,[R^{op}_{+-}]_{0,2}(\gamma_2+\gamma_0)...[R^{op}_{+-}]_{0,N-1}(\gamma_{N-1}+\gamma_0)\,[R^{op}_{+-}]_{0,N}(\gamma_N+\gamma_0),  
\end{aligned}
\end{equation}
where the notation $ [R_{++}]_{0,j}(\gamma_0-\gamma_j)$ and $ [R_{+-}^{ op}]_{0,j}(\gamma_j+\gamma_0)${{}, introduced in section \ref{sec:notation-and-conventions},} as usual indicates that the $R$-matrix acts in the spaces $0$ (the auxiliary space) and the $j$-th ``quantum" space. The reflection matrix $K$ (which coincides with the $K_-$ reflection of Sklyanin) acts in the auxiliary space only (with the wall itself being in the singlet representation of the symmetry algebra). 
Let us point out that throughout the whole paper we consider $N>1$, since for $N=1$ the Bethe ansatz trivialises.

The presence of the $R$-matrix $R_{+-}$
{{} is due} to the fact that in the double-row machinery the auxiliary particle ``coming back" after the reflection  scatters against all the quantum particles in reversed order. However in the massless case this means that the chirality of the auxiliary particle changes from right to left mover, in addition to a change of sign in the rapidity $\gamma_0$. As in the ordinary picture of the bulk algebraic Bethe ansatz, also in the double-row monodromy the quantum particles are considered to be instantaneously fixed: they are right- or left- movers (we have started with the case of all right-movers) but do not "move" really, meaning that there is no bulk $R$-matrix scattering two physical particles. The string of $R$-matrices appearing in the monodromy is always scattering the auxiliary particle (which is the only one really moving, going ``around") against the fixed pillars (the quantum particles) of the monodromy. The quantum particles' momenta are rather to be interpreted as impurities (``inhomogeneities"), that is, extra representation parameters sitting at each site of a spin-chain.

{{} As a sanity check we have explicitly verified} that $T_-$ in (\ref{Tminus}) solves the right-wall boundary Yang-Baxter equation (\ref{BYBE}) {{} by means of computer algebra} in the $N=2$ case.  To check this, one needs to introduce a second auxiliary space $0'$ and consider two copies of $T_-$, $T^0_-$ and $T^{0'}_-$, respectively acting in spaces $0$ (and quantum) and $0'$ (and quantum). The BYBE equation (\ref{BYBE}) reads therefore
\begin{eqnarray}
&&T^{0'}_-(\gamma_{0'}) [R^{op}_{+-}]_{0,0'}\Big(\gamma_{0'} - (-\gamma_0)\Big)T^0_-(\gamma_0)[R_{++}]_{0,0'}\Big( \gamma_0-\gamma_{0'}\Big) = \nonumber\\
&&\qquad \qquad [R^{op}_{--}]_{0,0'}\Big((-\gamma_{0'})-(-\gamma_0)\Big)T^0_-(\gamma_0)[R_{+-}]_{0,0'}\Big(\gamma_0 - (-\gamma_{0'})\Big)T_{-}^{0'}(\gamma_{0'}).
\label{BYBE2}
\end{eqnarray}
{{} In appendix \ref{BYBE2section}, we then provide the analytic proof for $N=1$ and the general proof follows by induction.}

To proceed, we need to find the correct way to write the {\it left} or {\it dual} equation of Sklyanin's \cite{Sklyanin:1988yz}. We should first clarify that its solution, denoted by $T_{+}$, represents the second building block of the full monodromy matrix $T\equiv T_{+}T_{-}$, where $T_{-}$ is the monodromy constructed above and solving the right-wall BYBE. In this sense, the new equation is {\it dual} to the right-wall BYBE, but this does not ensure a relation with the left-wall BYBE, like the ones studied in \cite{Bielli:2024xuv}, and for this reason we shall from now on refer to it as {\it dual} rather than {\it left}. 
{{} As mentioned in the Introduction, we will derive independently the dual equation tailored to our needs, and refer to the literature cited earlier for further discussion on the supersymmetric version.}


The simplest ansatz is to choose $T_+ = K_+$, with $K_+$ a suitable dual of the right-wall reflection matrix. In the singlet case this just means to postulate
\begin{eqnarray}
    T_+ = K_+(\gamma_0) = diag\big(1,g(\gamma_0)\big).\label{Vasi}
\end{eqnarray}
The function $g(\gamma_0)$ is then fixed by solving the dual equation. In appendix \ref{appendo} we will derive the dual equation and show that 
the full monodromy, traditionally obtained as \cite{Sklyanin:1988yz}
\begin{eqnarray}
    T(\gamma_0) = T_+(\gamma_0) T_-(\gamma_0),\label{commu}
\end{eqnarray}
satisfies the commutativity of the supertrace (transfer matrix) necessary for integrability:
\begin{eqnarray}
[\mbox{str}_0 T(\gamma_0),\mbox{str}_{0'} T(\gamma_{0'})]=0, \qquad \forall \gamma_0, \gamma_{0'}.\label{commur}
\end{eqnarray}



\noindent
{{} The dual equation is derived in appendix A} and reads 
\begin{eqnarray}
&&\Sigma_2 R^{op}_{--}(\gamma_-)\Sigma_2 \,  \big(T^1_+ (\gamma_1)\big)^{st_1} \, R_{+-}(\gamma_+) \, \big(T^2_+(\gamma_2)\big)^{(st_2)^3}  \nonumber\\
&&\qquad \qquad \qquad \qquad = \big(T^2_+(\gamma_2)\big)^{(st_2)^3} \, R^{op}_{+-}(\gamma_+) \, \big(T^1_+ (\gamma_1)\big)^{st_1} \, \Sigma_2 R_{++}(\gamma_-)\Sigma_2,\label{creq}
\end{eqnarray}
where
\begin{eqnarray}
    \gamma_\pm = \gamma_1 \pm \gamma_2,
\end{eqnarray}
and we recall that $\Sigma_2 = \mathbbmss{1} \otimes \sigma_3$, with $\sigma_3 = E_{11}-E_{22}$. We remind that the basis $E_{ij}$ is made of matrices with all zeroes but a $1$ in row $i$, column $j$ and the $R$-matrices $R_{++},R_{--},R_{+-}$ are explicitly given (ignoring dressing factors) {{} in section \ref{sec:notation-and-conventions}}. The symbol $(st_i)^3$ stands for the cube of the supertransposition, {\it i.e.} supertranposing three times, in the space $i$ ($i=1,2$). This is necessary because, in the supersymmetric case, it is only the fourth power of the supetransposition which returns the identity, {{}i.e. $(st)^3$ is the inverse of $st$}. We {{} stress} that {{} the dual equation, as we have chosen to derive it,} allows us to obtain commuting charges {{} even if} it does not explicitly feature the crossing transformation $\gamma_i \to \gamma_i + i \pi$. {{} It is also not (immediately, at least) related to the left-wall BYBE.}  

We can now find the most general solution to the dual equation (\ref{creq}) for the singlet boundary: we can simply adopt the diagonal ansatz (\ref{Vasi}) - which is the only option for a singlet wall - and notice that the supertranspose is immaterial on diagonal matrices. We therefore plug the ansatz in the equation and see that we can reduce it to a single condition linking $g(\gamma_1)$ and $g(\gamma_2)$. Imposing that there is no mixed dependence on the rapidities allows us to obtain a differential equation for each $g(\gamma_i)$ separately, whose solution depends on an integration constant $r \in \mathbbmss{C}$ and produces the result:
\begin{eqnarray}
g(\gamma_i) =  -i +\frac{2i(1+e^{\gamma_i})}{1+e^{\gamma_i} + e^{2 i r}(e^{\gamma_i}-1)}, \qquad i=1,2.\label{solu}   
\end{eqnarray}
Using the solution (\ref{solu}), (\ref{Vasi}), we have explicitly checked that {{}(\ref{commur})} holds for the case of two physical sites. We have also checked numerically that the eigenvectors of the transfer matrix do not depend on the auxiliary variable $\gamma_0$, which is a necessary condition for the commutativity for different values of the auxiliary variable. All this has been checked for arbitrary values of the integration constants $r$ and $c$: their assignment just distinguishes the amount of super-symmetry preserved by the boundary, as described in detail in \cite{Bielli:2024xuv}, and it does not affect the integrability (commutativity) property itself. We remind that the proof in this section is valid if we use the ansatz (\ref{Vasi}), since in this case $T_+$ does not act on the ``quantum" spaces and therefore we can swap $T^1_+$ and $T^2_-$ (which is, as in \cite{Sklyanin:1988yz}, one of the crucial assumptions of the proof of  commutativity). 


\subsection{Boundary Algebraic Bethe ansatz}

In this section we describe the procedure of the (boundary) algebraic Bethe ansatz for our {{} situation}. The idea is as follows: much as in the case of the ordinary (bulk) algebraic Bethe ansatz one relies on the so-called RTT relations, in this case we will rely on the right-wall BYBE \eqref{BYBE_original_form}, which rewritten in the form \eqref{BYBEa} can be called ``RTRT" relations. The spirit is the same: we write the monodromy matrix $T_-$ in terms of abstract operators $A,B,C,D$ 
\begin{eqnarray}
T_-(\gamma_0) = \begin{pmatrix}
A(\gamma_0)&B(\gamma_0)\\C(\gamma_0)&D(\gamma_0)\end{pmatrix}_0 = E_{11} \otimes A(\gamma_0) +E_{12} \otimes B(\gamma_0) +E_{21} \otimes C(\gamma_0) +E_{22} \otimes D(\gamma_0).  \label{tmi}
\end{eqnarray}
In this writing, the first space is the auxiliary $0$ and the rest of the spaces are the quantum spaces.

\subsubsection{{}One magnon case}

{{} To begin we focus on the one-magnon case and for simplicity, in this paragraph, whenever we provide explicit representations depending on the physical particles, we do so for $N=2$. In such case the two physical sites (particles) are denoted by spaces $1$ and $2$, on which the operators $A,B,C,D$ act}. 

We now introduce a second auxiliary space $0'$, so that we can write the right-wall boundary Yang-Baxter 
equation (which $T_-$ satisfies) in terms of these abstract elements $A,B,C,D$ - ordering the vector spaces as $V_0 \otimes V_{0'} \otimes V_1 \otimes V_2$. 
For instance, $T_-^0$ will be given by $E_{11} \otimes \mathbbmss{1} \otimes A(\gamma_0) + ...$, showing that nothing happens in space $0'$. Instead, $T_-^{0'}$ will be given by $\mathbbmss{1} \otimes E_{11} \otimes A(\gamma_{0'}) + ...$, showing that nothing happens in space $0$. The boundary Yang-Baxter equation (YBE) also features  four $R$-matrices, which act in the spaces $0$ and $0'$ only. The goal is to read the right boundary YBE as a set of commutation-type relations between $A(\gamma_0),B(\gamma_0),C(\gamma_0),D(\gamma_0)$ and $A(\gamma_{0'}),B(\gamma_{0'}),C(\gamma_{0'}),D(\gamma_{0'})$. These are all in principle non-commuting operators, since they have an action on the physical spaces. By substituing the actual entries of $T_-$, these RTRT relations will of course be identically satisfied (the actual entries of $T_-$ will in this way provide a {\it representation} of the abstract RTRT relations). 

We wrote down first by hand in components the RTRT relations, that is, the right boundary YBE with $T_-$ written in terms of abstract operators. By keeping track of all the fermionic signs, and setting 
\begin{eqnarray}
&&A(\gamma_0) \equiv  A_{11}(\gamma_0), \qquad B(\gamma_0) \equiv  A_{21}(\gamma_0), \qquad C(\gamma_0) \equiv  A_{12}(\gamma_0), \qquad D(\gamma_0) \equiv  A_{22}(\gamma_0), \nonumber \\
&&R_{\alpha\beta}(\gamma) =  r_{\alpha\beta}^{abcd}(\gamma) E_{ab} \otimes E_{cd}, \qquad \alpha,\beta = \pm,
\end{eqnarray}
we found, after some simplifications, that the right boundary Yang-Baxter equation (\ref{BYBE2}) becomes: 
{{}
\begin{align}
&r_{+-}^{abjd} (\!\gamma_{0'}\!+\!\gamma_0\!) \, r_{++}^{msdk} (\!\gamma_{0}\!-\!\gamma_{0'}\!) \, (\!-1\!)^{(|j|\!+\!|k|)(|i|\!+\!|j|)+(|b|\!+\!|s|)(|d|\!+\!|j|)+(|m|\!+\!|s|\!+\!|d|\!+\!|k|)(|m|\!+\!|b|)}  E_{as} \!\otimes\! E_{ik} \!\otimes\! A_{ji}(\!\gamma_{0'}\!) A_{mb}(\!\gamma_0\!) 
\notag \\
& \!=\! r_{--}^{aicd}(\!-\gamma_{0'}\!+\!\gamma_0\!) \, r_{+-}^{jndk}(\!\gamma_{0}\!+\!\gamma_{0'}\!) \, (\!-1\!)^{(|j|\!+\!|n|)(|c|\!+\!|d|\!+\!|i|\!+\!|j|)\!+\!(|c|\!+\!|q|)(|i|\!+\!|j|)} E_{an} \!\otimes\! E_{cq} \!\otimes\! A_{ji}(\!\gamma_0\!) A_{qk}(\!\gamma_{0'}\!) \,\, ,
\label{indices}
\end{align}}
where all the repeated indices are summed over.
The condition (\ref{indices}) amounts to 16 different relations, in correspondence with the 16 basis elements $E_{ab}\otimes E_{cd}$ which need to be matched on the left and the right hand side. Some of these 16 relations are just $0=0$ due to the six-vertex structure of the $R$-matrices. The list of 16 relations is {{} readily} obtained and we shall not spell it out explicitly here - it has the same content as (\ref{indices}). As an illustration, we report just 3 out of these 16 relations here below:
\begin{align}
& 0=-A(\gamma_0)A(\gamma_{0'}) + A(\gamma_{0'})A(\gamma_{0}) -i\Big[B(\gamma_{0})C(\gamma_{0'}) - B(\gamma_{0'})C(\gamma_{0})\Big]{\rm sech}\big(\frac{\gamma_0 + \gamma_{0'}}{2}\big), \notag \\
& \notag \\
& 0=-(\cosh \gamma_0  +  \cosh \gamma_{0'})A(\gamma_{0})B(\gamma_{0'}) +2\cosh\big(\frac{\gamma_0 + \gamma_{0'}}{2}\big)A(\gamma_{0'})B(\gamma_{0}) + 2 i \cosh\big(\frac{\gamma_0 - \gamma_{0'}}{2}\big)B(\gamma_{0})D(\gamma_{0'})
\notag \\
& \qquad \qquad \qquad \qquad + (\cosh \gamma_0  -  \cosh \gamma_{0'})B(\gamma_{0'})A(\gamma_{0})-2 i B(\gamma_{0'})D(\gamma_{0}), \notag \\
& \notag \\
& 0= B(\gamma_{0})B(\gamma_{0'})-B(\gamma_{0'})B(\gamma_{0}), \notag \\
&\notag \\
& \text{etc}...\label{disp}
\end{align}
Interestingly, the third relation in (\ref{disp}) is telling us that the $B$ operators commute among themselves: this means that creating eigenstates (magnons) by the recipe of creating excitations using $B$'s, as in
\begin{eqnarray}
|\mu_1,...,\mu_M\rangle \equiv B(\mu_1) B(\mu_2)...B(\mu_M) |0\rangle \,\, ,
\end{eqnarray}
where $|0\rangle$ is a pseudovacuum to be shortly defined (see also the Introduction),
is an unambiguous writing, because all the $B$'s mutually commute. {{} The proof of commutativity follows from the RTRT relations, realising that the only entries which could contribute two $B$ operators are schematically given by 
\begin{equation}
\begin{aligned}
& (1 \otimes E_{12} \otimes B')(- t_{\gamma_{0'} + \gamma_{0}} E_{11} \otimes E_{22} \otimes \mathbbmss{1} +E_{22} \otimes E_{22} \otimes \mathbbmss{1}+ i s_{\gamma_{0'} + \gamma_{0}} E_{12} \otimes E_{21} \otimes \mathbbmss{1})\times 
\\
&(E_{12} \otimes \mathbbmss{1} \otimes B)(E_{11} \otimes E_{11} \otimes \mathbbmss{1} -t_{\gamma_{0} - \gamma_{0'}} E_{11} \otimes E_{22} \otimes \mathbbmss{1} + s_{\gamma_{0} - \gamma_{0'}} E_{21} \otimes E_{12} \otimes \mathbbmss{1})=
\\
& = (E_{11} \otimes E_{11} \otimes \mathbbmss{1} -t_{\gamma_{0} - \gamma_{0'}} E_{22} \otimes E_{11} \otimes \mathbbmss{1} + i s_{\gamma_{0} - \gamma_{0'}} E_{12} \otimes E_{21} \otimes \mathbbmss{1})(E_{12} \otimes \mathbbmss{1} \otimes B) 
\times 
\\
& \quad\,\, (t_{\gamma_{0} + \gamma_{0'}} E_{22} \otimes E_{11} \otimes \mathbbmss{1} -E_{22} \otimes E_{22} \otimes \mathbbmss{1} + s_{\gamma_{0} + \gamma_{0'}} E_{21} \otimes E_{12} \otimes \mathbbmss{1})(1 \otimes E_{12} \otimes B'),
\end{aligned}
\end{equation}
where we defined $t_x \equiv \tanh \frac{x}{2}$ and $s_x \equiv \mbox{sech}\frac{x}{2}$.
Explicit (graded) multiplication of the terms leads to
\begin{eqnarray}
t_{\gamma_0 + \gamma_{0'}} E_{12}\otimes E_{12} \otimes B' B =   t_{\gamma_0 + \gamma_{0'}} E_{12}\otimes E_{12} \otimes B B',  
\end{eqnarray}
which proves the statement. We recall that $B$ is fermionic, so swapping a $B$ and $E_{12}$ or $E_{21}$ introduces a minus sign. This does not prevent the $B$s from commuting\footnote{{}One example of fermionic matrix which commutes for different values of the argument (and is also not nilpotent) is 
\begin{eqnarray}
m(t)= \frac{1-t(4+t)}{t^2-1} E_{11}\otimes E_{21} + \frac{ 1+t(4-t)}{t^2-1} E_{22}\otimes E_{21} +   E_{21}\otimes E_{11} + E_{21}\otimes E_{22} \,\, .
\end{eqnarray} 
}.
}

{{}The} relations (\ref{indices}) and their breaking down in components are {{} valid} for generic $N$ in (\ref{Tminus}), because they are valid for any $T_-$ as long as the latter satisfies the right boundary YBE.

A suitable pseudovacuum is easily found: using (\ref{tmi}) and recalling the definition \eqref{commu} together with the choice \eqref{Vasi}, one can convince oneself that the complete double row transfer matrix $\mbox{str}_0 T(\gamma_0) = \mbox{str}_0\Big[T_+(\gamma_0)T_-(\gamma_0)\Big]$ can be written as 
\begin{eqnarray}
  \mbox{str}_0 T(\gamma_0) = A(\gamma_0) - g(\gamma_0) D(\gamma_0),  
\end{eqnarray}
where $g(\gamma_0)$ is defined in (\ref{solu}). We also explicitly obtain the operators $A(\gamma_0), B(\gamma_0),C(\gamma_0),D(\gamma_0)$ as matrices acting on two physical particles, because we know (\ref{Tminus}). {{}We} can therefore immediately verify that the state 
\begin{eqnarray}
    |0\rangle \equiv |\phi\rangle \otimes |\phi\rangle
\end{eqnarray}
is an adequate pseudovacuum, satisfying all the expected properties for such a state \cite{Fedor}:
\begin{eqnarray}
A(\gamma_0)|0\rangle = \lambda_1(\gamma_0)|0\rangle, \qquad D(\gamma_0)|0\rangle = \lambda_2(\gamma_0)|0\rangle, \qquad C(\gamma_0) |0\rangle = 0,\qquad B(\gamma_0) |0\rangle \neq 0, \label{ned} 
\end{eqnarray}
where we have left implicit the dependence on the physical rapidities. {{} In the case $N=2$, for example,} $\lambda_1$ and $\lambda_2$ are given by\footnote{As we said at the beginning, we have ignored the dressing factors for the moment - it is very easy to reinstate them at the end as an overall factor. }
\begin{align}
&\lambda_1 (\gamma_0) \!=\! 1, \label{lambda} \\
&\lambda_2 (\gamma_0) \!=\!  -i\Biggl[\frac{2}{\cosh \! \gamma_0\!  \!+\! \cosh \!\gamma_2\!} \!+\! \frac{\tanh \!\big(\frac{\gamma_0 - \gamma_2}{2}\big)\! \tanh \!\big(\frac{\gamma_0 + \gamma_2}{2}\big)\! \Bigl( [\cosh \!\gamma_0\! \!+\! \cosh \!\gamma_1\! ]\tanh[{\rm arccoth} \, e^{\gamma_0} \!+\! i c^{L}] \!-\! 2\Bigr)}{\cosh \!\gamma_0\! \!+\! \cosh \!\gamma_1\! }\Biggr]. 
\notag
\end{align}  
The transfer matrix is immediately diagonal on the pseudovacuum with eigenvalue
\begin{eqnarray}
\mbox{str}_0 T(\gamma_0)|0\rangle = \Big[\lambda_1(\gamma_0) - g(\gamma_0) \lambda_2(\gamma_0)\Big]|0\rangle.    
\end{eqnarray}
The real challenge is to figure out whether the one-magnon state
\begin{eqnarray}
|\mu_1\rangle = B(\mu_1)|0\rangle    
\end{eqnarray}
is an eigenstate of the transfer matrix. For this, we proceeded as follows: we stared at the set of 16 RTRT relations, until we isolated those relations which involve a $B$ operator preceded by either an $A$ or a $D$ operator. There are several of such relations. We did put all these particular relations together, and manipulated them so as to recast them in the most useful way. This involved taking linear combinations of the original relations, and of the relations obtained exchanging $\gamma_0$ and $\gamma_{0'}$, to obtain further relations. We managed to reduce them to the following two relations:
\begin{eqnarray}
&&A(\gamma_0) B(\gamma_{0'}) = \frac{1}{2} {\rm sech}\big(\frac{\gamma_0 + \gamma_{0'}}{2}\big) \Big(2 i B(\gamma_0) D(\gamma_{0'}) + {\rm csch}\big(\frac{\gamma_0 - \gamma_{0'}}{2}\big)\big[B(\gamma_{0'}) A(\gamma_{0})(\sinh \gamma_0 + \sinh \gamma_{0'}) \nonumber\\
&& \qquad \qquad \qquad -2B(\gamma_{0}) A(\gamma_{0'}) \,{\rm sinh}\big(\frac{\gamma_0 + \gamma_{0'}}{2}\big) \big]\Big),\nonumber\\
&&D(\gamma_0) B(\gamma_{0'}) = \frac{1}{2} {\rm sech}\big(\frac{\gamma_0 + \gamma_{0'}}{2}\big) \Big(-2 i B(\gamma_0) A(\gamma_{0'}) + {\rm csch}\big(\frac{\gamma_0 - \gamma_{0'}}{2}\big)\big[B(\gamma_{0'}) D(\gamma_{0})(\sinh \gamma_0 + \sinh \gamma_{0'}) \nonumber\\
&& \qquad \qquad \qquad -2B(\gamma_{0}) D(\gamma_{0'}) \,{\rm sinh}\big(\frac{\gamma_0 + \gamma_{0'}}{2}\big) \big]\Big).\label{use}
\end{eqnarray}
At this point, the game is won. In fact, we simply say that the transfer matrix acting on the one-magnon state leads to
\begin{eqnarray}
\Big[\mbox{str}_0 T(\gamma_0)\Big]B(\mu_1)|0\rangle = A(\gamma_0)B(\mu_1)|0\rangle - g(\gamma_0) D(\gamma_0)B(\mu_1)|0\rangle. 
\end{eqnarray}
But now we can use (\ref{use}) to commute the $A$ and $D$ operators through the $B$ operator. We get
\begin{equation}
\label{rew}
\begin{aligned}
&\Big[\mbox{str}_0 T(\gamma_0)\Big]B(\mu_1)|0\rangle =
\\
&\qquad \frac{1}{2} {\rm sech}\big(\frac{\gamma_0 + \mu_1}{2}\big) \Big(2 i B(\gamma_0) D(\mu_1)|0\rangle + {\rm csch}\big(\frac{\gamma_0 - \mu_1}{2}\big)\big[B(\mu_1) A(\gamma_{0})|0\rangle(\sinh \gamma_0 + \sinh \mu_1) 
\\
& \quad  -2B(\gamma_{0}) A(\mu_1)|0\rangle \,{\rm sinh}\big(\frac{\gamma_0 + \mu_1}{2}\big) \big]\Big) - g(\gamma_0) \Bigg(\frac{1}{2} {\rm sech}\big(\frac{\gamma_0 + \mu_1}{2}\big) \Big(-2 i B(\gamma_0) A(\mu_1)|0\rangle 
\\
& \quad + {\rm csch}\big(\frac{\gamma_0 - \mu_1}{2}\big)\big[B(\mu_1) D(\gamma_{0})|0\rangle(\sinh \gamma_0 + \sinh \mu_1)    -2B(\gamma_{0}) D(\mu_1)|0\rangle \,{\rm sinh}\big(\frac{\gamma_0 + \mu_1}{2}\big) \big]\Big)\Bigg).
\end{aligned}
\end{equation}
But we also know that the $A$ and $D$ operators act diagonally on the pseudovacuum thanks to (\ref{ned}), and so we can everywhere just substitute the eigenvalues:
\begin{equation}
\label{somer}
\begin{aligned}
&\Big[\mbox{str}_0 T(\gamma_0)\Big]B(\mu_1)|0\rangle = 
\\
&\qquad \frac{1}{2} {\rm sech}\big(\frac{\gamma_0 + \mu_1}{2}\big) \Big(2 i B(\gamma_0) \lambda_2(\mu_1)|0\rangle + {\rm csch}\big(\frac{\gamma_0 - \mu_1}{2}\big)\big[B(\mu_1) \lambda_1(\gamma_{0})|0\rangle(\sinh \gamma_0 + \sinh \mu_1) 
\\
& \quad  -2B(\gamma_{0}) \lambda_1(\mu_1)|0\rangle \,{\rm sinh}\big(\frac{\gamma_0 + \mu_1}{2}\big) \big]\Big) - g(\gamma_0) \Bigg(\frac{1}{2} {\rm sech}\big(\frac{\gamma_0 + \mu_1}{2}\big) \Big(-2 i B(\gamma_0) \lambda_1(\mu_1)|0\rangle 
\\
& \quad +{\rm csch}\big(\frac{\gamma_0 - \mu_1}{2}\big)\big[B(\mu_1) \lambda_2(\gamma_{0})|0\rangle(\sinh \gamma_0 + \sinh \mu_1)    -2B(\gamma_{0}) \lambda_2(\mu_1)|0\rangle \,{\rm sinh}\big(\frac{\gamma_0 + \mu_1}{2}\big) \big]\Big)\Bigg).
\end{aligned}
\end{equation}
This is not yet showing that the one-magnon state is an eigenstate: some terms in (\ref{somer}) are good, meaning, they reconstruct the one magnon state $B(\mu_1)|0\rangle$. Other terms are spurious (unwanted), because they feature $B(\gamma_0)|0\rangle$. But this is no problem: we set these unwanted terms to zero and read off the Bethe equations. By collecting these spurious terms we find that we need to impose
\begin{equation}
\label{rela}
\begin{aligned}
&  \frac{1}{2} {\rm sech}\big(\frac{\gamma_0 + \mu_1}{2}\big) \Big(2 i  \lambda_2(\mu_1) + {\rm csch}\big(\frac{\gamma_0 - \mu_1}{2}\big)\big[   -2 \lambda_1(\mu_1) \,{\rm sinh}\big(\frac{\gamma_0 + \mu_1}{2}\big) \big]\Big)
\\
& - g(\gamma_0) \Bigg(\frac{1}{2} {\rm sech}\big(\frac{\gamma_0 + \mu_1}{2}\big) \Big(-2 i  \lambda_1(\mu_1) + {\rm csch}\big(\frac{\gamma_0 - \mu_1}{2}\big)\big[-2 \lambda_2(\mu_1) \,{\rm sinh}\big(\frac{\gamma_0 + \mu_1}{2}\big) \big]\Big)\Bigg)=0 \,\, ,
\end{aligned}
\end{equation}
since all the unwanted terms have a $B(\gamma_0)|0\rangle$ in common and no other operators. By using the explicit expression for $g(\gamma_0)$ in (\ref{solu}) one can see that (\ref{rela}) reduces to 
\begin{eqnarray}
\label{Bethe-simplified}
e^{2i r} \big( \lambda_2(\mu_1)-i\big) \sinh \frac{\mu_1}{2} - \big( \lambda_2(\mu_1)+i\big) \cosh \frac{\mu_1}{2} =0,   
\end{eqnarray}
having also used $\lambda_1(\gamma)=1$, with $r$ being the integration constant in (\ref{solu}), and where the dependence on $\gamma_0$ is rightfully disappeared - in accordance with the fact that integrability dictates that the eigenvectors should not depend on the auxiliary variable $\gamma_0$.
{{}Assuming that} we know the function $\lambda_2(\mu_1)$ {{} for generic $N$} from (\ref{lambda}), 
we see that (\ref{Bethe-simplified}) is the {{} Bethe} equation linking the one-magnon rapidity $\mu_1$, to use in $B(\mu_1)|0\rangle$ to get a true eigenvector, with the inhomogeneites {{}$\gamma_i$, $i=1,...,N$}. Since the unwanted terms are now cancelled, we can focus on the good terms in $(\ref{somer})$ and read off the transfer-matrix eigenvalue of the one-magnon state:
\begin{align}
\Big[\mbox{str}_0 T(\gamma_0)\Big]B(\mu_1) &|0\rangle  = \Bigg(\frac{1}{2} {\rm sech}\big(\frac{\gamma_0 + \mu_1}{2}\big)  {\rm csch}\big(\frac{\gamma_0 - \mu_1}{2}\big) \lambda_1(\gamma_{0})(\sinh \gamma_0 + \sinh \mu_1) 
\label{frome} \\
& \qquad    - g(\gamma_0) \frac{1}{2} {\rm sech}\big(\frac{\gamma_0 + \mu_1}{2}\big)   {\rm csch}\big(\frac{\gamma_0 - \mu_1}{2}\big) \lambda_2(\gamma_{0})(\sinh \gamma_0 + \sinh \mu_1)     \Bigg)B(\mu_1)|0\rangle \,\, ,
\notag
\end{align}
where the rapidities are subject to the Bethe equation (\ref{Bethe-simplified}), which can compactly be rewritten as
\begin{eqnarray}
\label{Bethe}
e^{2ir} \frac{\big(\lambda_2(\mu_1)-i\big)}{\big( \lambda_2(\mu_1)+i\big)} \, \tanh \frac{\mu_1}{2}=1.
\end{eqnarray}
Let us remark that the formula (\ref{Bethe}), when written with  $\lambda_2$ left implicit, is {{} the} general Bethe equation for one-magnon states, for any number of physical excitations $N$. This is because to obtain it we have just used the RTRT relations, which are general. The limitation to two sites comes in when we wish to explicitly insert the values of  $\lambda_2$, since so far we only have (\ref{lambda}) - the two-site expression\footnote{With our conventions it is easy to prove that $\lambda_1$ is equal to $1$ for any $N$ (see appendix \ref{B}), so the stumbling block is really only to compute the general expression for $\lambda_2$.}. 
{{}In the following sections we will construct the general expression for any $N$ and we will prove its validity analytically.}

\subsubsection{General case}
With the help of {{} computer algebra}, we have iterated the procedure of the previous subsection in the case of two magnons:
\begin{eqnarray}
    |\mu_1,\mu_2\rangle \equiv B(\mu_1) B(\mu_2)|0\rangle.
\end{eqnarray}
Taking advantage of the fact that the $B$ operators mutually commute, {{} a compact result can be reached}. Schematically, one finds
\begin{eqnarray}
\Big[{\rm str_0} T(\gamma_0)\Big] B(\mu_1) B(\mu_2)|0\rangle = \Lambda_2 B(\mu_1) B(\mu_2)|0\rangle + U_1 B(\mu_1) B(\gamma_0)|0\rangle + U_2 B(\gamma_0) B(\mu_2)|0\rangle,   
\end{eqnarray}
where the functions $\Lambda_{2},U_1,U_2$ are explicitly produced by Mathematica. The first term, involving $\Lambda_2$, is the eigenvalue, while the other two terms, involving $U_1$ and $U_2$, respectively, are both unwanted terms. We study these terms separately.

The eigenvalue $\Lambda_2$, whose expression is rather complicated, is best written in terms of $\lambda_2$ from (\ref{lambda}) left implicit. $\Lambda_2$ is the two-magnon eigenvalue, and it has to be compared with the one-magnon eigenvalue read-off from (\ref{frome}) - which we can call $\Lambda_1$. We stare at the two expressions for $\Lambda_2$
and $\Lambda_1$, and we can come up with an {{} expression} for how the formula proceeds for any $M$ number of magnons:
\begin{eqnarray}
\Lambda_M = -\frac{1}{2^M} \Big[g(\gamma_0) \lambda_2(\gamma_0) - 1\Big]\prod_{i=1}^M \Big(\sinh \gamma_0 + \sinh \mu_i\Big)\, {\rm csch} \big(\frac{\gamma_0 - \mu_i}{2}\big) \, {\rm sech }\big(\frac{\gamma_0 + \mu_i}{2}\big), \label{generale}
\end{eqnarray}
with the understanding that the product in (\ref{generale}) is equal to $1$ for $M=0$ (no magnons, that is the pseudovacuum).
{{}With} $\lambda_2$ left implicit, the expression (\ref{generale}) is valid for any {{} $M$ and $N$ - one} needs to substitute the appropriate value of $\lambda_2$ for $N$ sites. {{} In appendix \ref{gene} we provide an analytic proof of the formula (\ref{generale}) using a recursion on $N$.}

Similarly, we can treat the unwanted terms. We compare the unwanted term in the previous section for the one magnon state, formula (\ref{rela}), with the two-magnon expressions $U_1$ and $U_2$. 
A quick inspection reveals that these formulas suggest a clear generalisation for a generic number of magnons $M$:
\begin{eqnarray}
i \big(g(\gamma_0) + \lambda_2 (\mu_j) \big) + {\rm csch } \big(\frac{\gamma_0 - \mu_j}{2}\big) \, {\rm sinh} \big(\frac{\gamma_0 + \mu_j}{2}\big) \, \big(-1 + g(\gamma_0)\lambda_2(\mu_j)\big)=0, \qquad j=1,...,M.\label{right}
\end{eqnarray}
This is again a {{} formula} valid for any number of sites (physical particles) as long as we write the formula in terms of the quantity $\lambda_2$.  {{} In appendix \ref{gene} we provide a proof of this fact as well.} We can therefore obtain the Bethe equations of the problem by simplifying (\ref{right}) using the explicit form of $g(\gamma_0)$ in (\ref{solu})
\begin{eqnarray}
\label{Bethe2}
e^{2ir} \frac{\big( \lambda_2(\mu_j)-i\big)}{\big( \lambda_2(\mu_j)+i\big)} \, \tanh \frac{\mu_j}{2}=1, \qquad j=1,...,M,
\end{eqnarray}
where again the dependence on $\gamma_0$ rightfully disappears. This formula is now valid for any $M$ and $N$, and for $j=1$ it clearly reduces to the one-magnon case previously encountered. The way to construct {{} the $2^N$} eigenstates is then as usual to list all the solutions for $\mu$ of the equation $e^{2ir} \frac{\big( \lambda_2(\mu)-i\big)}{\big( \lambda_2(\mu)+i\big)} \, \tanh \frac{\mu}{2}=1$, and to {{} plug into} $B(\mu_1)...B(\mu_k)|0\rangle$, for $k=0,...,M$. An example of this can be found in appendix B of \cite{DiegoBogdanAle} {{} - in that case the inequivalent solutions to the Bethe equations are exactly $N$, so that choosing $M$ distinct roots out of $N$ for a $M$-magnon state produces a total of $\sum_{M=0}^N {{N}\choose{M}} = 2^N$ as expected for a quantum space ${\big[\mathbbmss{C}^{(1|1)}\big]}^N$. We have not performed a thoroughly enough numerical study of the solutions of (\ref{Bethe2}) to be able to list the eigenvectors as in \cite{DiegoBogdanAle} - we reserve a full analysis for future work.}

We conclude by reporting the following verification which we have made. Suppose for a moment to construct the algebraic Bethe ansatz as we have done, but without knowing in full details the structure of the dual equation \eqref{creq}, or equivalently, without knowing the explicit form of the $T_+$ monodromy except that it must be diagonal for singlet boundaries. We can always normalise $T_+$ with a $1$ in the boson-boson entry by choosing $T_+$ to be the regular solution of the dual reflection boundary Yang-Baxter equation with trivial action on the quantum spaces, and rescaling the dressing factor (omitted for simplicity in the above analysis). Therefore, we could have performed the whole procedure with $g(\gamma_0)$ left unknown. If one  considers then the unwanted term (\ref{right}), one could try to {\it solve} for $g(\gamma_0)$ in such a way that the dependence of $\gamma_0$ in (\ref{right}) effectively disappears - {\it i.e.} this dependence is accumulated only in an overall factor, which drops out when setting the unwanted term to $0$. This is also equivalent to imposing that, solving (\ref{right}) for $\lambda_2$ in terms of $g(\gamma_0)$, then $g(\gamma_0)$ is such that the solution for $\lambda_2$ is independent of $\gamma_0$. If we do that, we find that the  {\it only} solution compatible with this integrability-dictated request is precisely (\ref{solu}). This confirms that the dual equation (\ref{creq}) which we have found, at least in the singlet case, produces the most general solution for $T_+=K_+$ compatible with integrability, given our form for $T_-$.   
  
\subsubsection{The general $\lambda_2$ \label{general}}

The last remaining task to declare the algebraic Bethe ansatz complete is to find the general expression for the vacuum-eigenvalue of the $D$ operator, called $\lambda_2$. {{} We have first pushed the brute force calculation} to $N=3$, and constructed the associated operators $A,B,C,D$. The pseudovacuum is as usual 
\begin{eqnarray}
    |0\rangle_3 = |\phi\rangle \otimes |\phi\rangle \otimes |\phi\rangle.
\end{eqnarray}
The operator $A$ acts, exactly as we have proven in general, as $A|0\rangle_3 = |0\rangle_3$, while $D$ acts (diagonally) with an eigenvalue which is quite complicated. However, staring at the expression produced by the brute-force computation for three sites, and at the expression for two sites given by (\ref{lambda}), we spotted a conjecture for the general trend for any number of sites $N$. It looks like this:
\begin{align}
\lambda_2(\gamma_0;\gamma_1,...,\gamma_N)  \!=\! & -\frac{2i}{\cosh{\!\gamma_{0}\!}\!+\!\cosh{\!\gamma_{N}\!}}\!-\! \sum_{n=1}^{N-1} \Biggl[  \frac{2i(-1)^{n}}{\cosh{\!\gamma_{0}\!}\!+\!\cosh{\!\gamma_{N\!-\!n}\!}} \, \prod_{k=N\!-\!n\!+\!1}^{N} \tanh{\!\frac{\gamma_{0}\!-\!\gamma_{k}}{2}\!}\tanh{\!\frac{\gamma_{0}\!+\!\gamma_{k}}{2}\!} \Biggr]+
\notag \\
& - (-1)^N \frac{i(2 \!+\! \xi \!+\! \xi \, e^{\gamma_0})}{(\xi +(2+\xi)  e^{\gamma_0})} \prod_{k=1}^{N} \tanh{\!\frac{\gamma_{0}\!-\!\gamma_{k}}{2}\!}\tanh{\!\frac{\gamma_{0}\!+\!\gamma_{k}}{2}\!}, \label{exal}
\end{align}

where 
\begin{eqnarray}
    \xi = e^{2i \, c^L}-1.
\end{eqnarray}
{{} We have first explicitly checked that this formula holds up to $N=5$ and then achieved a proof using induction. For this, we schematically write the case for $N+1$ sites in the following way:
\begin{eqnarray}
T_-^{(N+1)} = [E_{kl} \otimes \mathbbmss{1}_N \otimes \mu_{lk}][E_{ij}\otimes A_{ji} \otimes \mathbbmss{1}][E_{mn} \otimes \mathbbmss{1}_N \otimes \nu_{nm}],\label{ex}  
\end{eqnarray}
where $\mathbbmss{1}_N$ denotes the N-fold tensor product of identities $\mathbbmss{1} \otimes ... \otimes \mathbbmss{1}$ and
\begin{eqnarray}
    E_{kl} \otimes \mu_{lk} = R_{++}, \qquad E_{mn}  \otimes \nu_{nm} = R^{op}_{+-}, \qquad T_-^{(N)} = E_{ij}\otimes A_{ji} \,\, .
\end{eqnarray}
The spaces in (\ref{ex}) are ordered as usual as $0-1-...-N-(N+1)$ and repeated indices are summed over. Inspection of the two matrices $R_{++}$ and $R^{op}_{+-}$ reveals that
\begin{equation}
\begin{aligned}
\mu_{11} &= E_{11} - \tanh \Big(\frac{\gamma_0 - \gamma_{N+1}}{2}\Big) E_{22}, 
\qquad \qquad 
\mu_{22} = - E_{22} + \tanh \Big(\frac{\gamma_0 - \gamma_{N+1}}{2}\Big) E_{11}, 
\\
\mu_{12} &= \mbox{sech} \Big(\frac{\gamma_0 - \gamma_{N+1}}{2}\Big) E_{12}, 
\qquad \qquad \qquad \quad \,
\mu_{21} = -\mbox{sech} \Big(\frac{\gamma_0 - \gamma_{N+1}}{2}\Big) E_{21},
\\
\nu_{11} &= E_{11} - \tanh \Big(\frac{\gamma_0 + \gamma_{N+1}}{2}\Big) E_{22}, 
\qquad \qquad \, 
\nu_{22} =  E_{22} - \tanh \Big(\frac{\gamma_0 + \gamma_{N+1}}{2}\Big) E_{11}, 
\\
\nu_{12} &= - i \, \mbox{sech} \Big(\frac{\gamma_0 + \gamma_{N+1}}{2}\Big) E_{12}, 
\qquad \qquad \qquad
\nu_{21} = i \, \mbox{sech} \Big(\frac{\gamma_0 + \gamma_{N+1}}{2}\Big) E_{21}.
\end{aligned}
\end{equation}
We can then explicitly simplify (\ref{ex}) and obtain
\begin{eqnarray}
T_-^{(N+1)} = (-1)^{(|j|+|n|)(|j|+|k|)} E_{kn} \otimes A_{ji} \otimes \mu_{ik} \, \nu_{nj} \,\, ,    
\end{eqnarray}
such that focusing on the $E_{22}$ entry to extract the operator $D$, we have
\begin{eqnarray}
D^{(N+1)} = (-1)^{(|j|+1)}  A_{ji} \otimes \mu_{i2} \, \nu_{2j} \,\, ,  
\end{eqnarray}
which can be interpreted as a coproduct map for the operator $D$ (see also \cite{Doikou:2004qd}).
We now remember that all we need to do is acting on the pseudovacuum, and hence compute
\begin{eqnarray}
D^{(N+1)}|0\rangle_{N+1} = (-1)^{(|j|+1)}  A_{ji} |0\rangle_N \otimes \mu_{i2} \, \nu_{2j} |\phi\rangle.  
\end{eqnarray}
Fortunately, we can use by the induction hypothesis that 
\begin{eqnarray}
C^{(N)}|0\rangle_N =A_{12}|0\rangle_N , \qquad  A^{(N)}|0\rangle_N =A_{11}|0\rangle_N =|0\rangle_N, \qquad D^{(N)}|0\rangle_N =A_{22}|0\rangle_N = \lambda_2^{(N)} |0\rangle_N \,\, , 
\end{eqnarray}
and by direct computation we have
\begin{equation}
\begin{aligned}
\mu_{12} \nu_{21} |\phi\rangle & = i   \mbox{sech} \Big(\frac{\gamma_0 - \gamma_{N+1}}{2}\Big) \mbox{sech} \Big(\frac{\gamma_0 + \gamma_{N+1}}{2}\Big) |\phi\rangle,
\\
\mu_{22} \nu_{22} |\phi\rangle & =    -\mbox{tanh} \Big(\frac{\gamma_0 - \gamma_{N+1}}{2}\Big) \mbox{tanh} \Big(\frac{\gamma_0 + \gamma_{N+1}}{2}\Big) |\phi\rangle,
\\
\mu_{12} \nu_{22} |\phi\rangle  & = 0 \,\, .\label{exac}
\end{aligned}
\end{equation}
Therefore (using again $|0\rangle_{N+1} = |0\rangle_N \otimes |\phi\rangle$) we obtain that the only terms contributing are 
\begin{eqnarray}
D^{(N+1)} |0\rangle_{N+1} = - A^{(N)} |0\rangle_N \otimes \mu_{12} \nu_{21} |\phi\rangle + D^{(N)}  |0\rangle_N \otimes \mu_{22} \nu_{22} |\phi\rangle \,\, ,  
\end{eqnarray}
and obtain the recursion relation
\begin{eqnarray}
\lambda_2^{(N+1)} = - i \,  \mbox{sech} \Big(\frac{\gamma_0 - \gamma_{N+1}}{2}\Big) \mbox{sech} \Big(\frac{\gamma_0 + \gamma_{N+1}}{2}\Big) -\mbox{tanh} \Big(\frac{\gamma_0 - \gamma_{N+1}}{2}\Big) \mbox{tanh} \Big(\frac{\gamma_0 + \gamma_{N+1}}{2}\Big) \, \lambda_2^{(N)} \,\, .\label{reco}
\end{eqnarray}
This is precisely solved by our expression (\ref{exal}). In fact, we can write it in a skeleton form as
\begin{eqnarray}
 x_N = s_N + \sum_{n=1}^{N-1} (-1)^n s_{N-n} \prod_{k=N-n+1}^N t_k - (-1)^N c_0 \prod_{k=1}^N t_k, 
\end{eqnarray}
where $x_k=T_-^{(k)}$, $c_0 = \frac{i(2+\xi +\xi e^{\gamma_0})}{\xi + (2+\xi)e^{\gamma_0}}$,  $s_k = - i \,  \mbox{sech} \Big[\frac{\gamma_0 - \gamma_{k}}{2}\Big] \mbox{sech} \Big[\frac{\gamma_0 + \gamma_{k}}{2}\Big]$ and $t_k = \mbox{tanh} \Big[\frac{\gamma_0 - \gamma_{k}}{2}\Big] \mbox{tanh} \Big[\frac{\gamma_0 + \gamma_{k}}{2}\Big]$.
Therefore it is easy to compute the skeleton form of $x_{N+1}$: changing summation variable as $n-1=m$ to readjust the sum, we can achieve
\begin{equation}
\begin{aligned}
x_{N+1} &= s_{N+1} + \sum_{m=0}^{N-1}(-1)^{m+1} s_{N-m} \prod_{k=N-m+1}^{N+1} t_k + (-1)^N c_0 \prod_{k=1}^{N+1} t_k =
\\
&= s_{N+1} - s_N \, t_{N+1} - \sum_{m=1}^{N-1}(-1)^{m} s_{N-m} \prod_{k=N-m+1}^{N+1} t_k  + (-1)^N c_0\prod_{k=1}^{N+1} t_k =
\\
& = s_{N+1} -  x_N t_{N+1} \,\, ,
\end{aligned}
\end{equation}
but this is exactly of the same form as the r.h.s. of (\ref{reco}).
}

\subsubsection{Left {\it vs.} right quantum particles}

We have repeated the calculation in the case where some of the quantum spaces (quantum particles) have left-moving chirality. The auxiliary particle is kept always to be right-moving - that should be sufficient to construct eigenstates for any configuration of inhomogeneities. We have performed {{} checks} for $N=2$ in all possible configurations/representations. This means that, in addition to the right-right case considered earlier, we have also tested the following three representations for $T_-$: mixed right-left, and both left, that is
\begin{equation}
\label{defu}
\begin{aligned}
&[R_{+-}]_{0,2}(\gamma_0\!-\!\gamma_2)\,[R_{++}]_{0,1}(\gamma_0\!-\!\gamma_1)K_{0}(\gamma_0) \, [R^{op}_{+-}]_{0,1}(\gamma_1\!+\!\gamma_0)\,[R^{op}_{--}]_{0,2}(\gamma_2\!+\!\gamma_0), \quad \gamma_1 <0,\gamma_2>0,
\\ 
&[R_{++}]_{0,2}(\gamma_0 \!-\! \gamma_2)\,[R_{+-}]_{0,1}(\gamma_0 \!-\! \gamma_1)K_{0}(\gamma_0) \, [R^{op}_{--}]_{0,1}(\gamma_1 \!+\! \gamma_0)\,[R^{op}_{+-}]_{0,2}(\gamma_2 \!+\! \gamma_0), \quad \gamma_1 >0,\gamma_2<0,
\\ 
&[R_{+-}]_{0,2}(\gamma_0 \!-\! \gamma_2)\,[R_{+-}]_{0,1}(\gamma_0 \!-\! \gamma_1)K_{0}(\gamma_0) \, [R^{op}_{--}]_{0,1}(\gamma_1 \!+\! \gamma_0)\,[R^{op}_{--}]_{0,2}(\gamma_2 \!+\! \gamma_0),\quad \gamma_1 >0,\gamma_2>0. 
\end{aligned}
\end{equation}
Because we have checked that these all satisfy the same boundary Yang-Baxter equation (\ref{BYBE2}), the RTRT relations are the same as we have found before (the representation of the operators $A,B,C,D$ being different). The $T_+$ solution can be taken the same as before, using which we have verified the commutativity of the transfer matrices for different auxiliary parameters in all cases. The Bethe equations are the same as long as we substitute the appropriate $\lambda_2$ {{} function}.

To our surprise, in all cases the function $\lambda_2$ is exactly the same function that we have already found, formula (\ref{lambda}). We consider this occurrence quite remarkable, and we have not yet found a {{} deeper} theoretical explanation for this fortunate coincidence. {{} We} have checked that this coincidence persists up to $N=5$: by extending in the obvious way the definitons (\ref{defu}) and repeating the calculation in each case, the expression for $\lambda_2$ is always the same for all possible different configuration of quantum particles, as long as the auxiliary particle is always a right mover. This is particularly convenient, because it simply allows to extend the Bethe equations to $\gamma_i \in \mathbbmss{R}, i=1,..,N$, and encompass in one go every possible combination of positive (right-mover) and negative (left-mover) chiralities, despite the fact that representations for right and left movers are different.  

{{} A practical explanation for this is given by the following fact\footnote{We thank the anonymous referee for the suggestion.}. In the appropriate domain one can verify that there are the following relations between our $R$-matrices:
\begin{equation}\label{what}
\begin{aligned}
R_{++}(\gamma) &= U R_{-+}(\gamma) U = V R_{+-}(\gamma) V,
\qquad  
R_{+-}(\gamma) = U R_{--}(\gamma) U, 
\qquad 
R_{-+}(\gamma)= V R_{--}(\gamma) V,  
\\
R^{op}_{--} &= V R^{op}_{+-}V,
\end{aligned}
\end{equation}
where $U = \mathbbmss{1} \otimes \mbox{diag}(1,-i)$ and $V = \mbox{diag}(1,i) \otimes \mathbbmss{1}$. 
We take for instance the first transfer matrix in (\ref{defu}), which can can be schematically rewritten as
\begin{eqnarray}
V_{02}^{-1} [R_{++}]_{02} V_{02}^{-1} [R_{++}]_{01} K_0 [R^{op}_{+-}]_{01} V_{02} [R^{op}_{+-}]_{02} V_{02}. 
\end{eqnarray}
At this point it is possible to verify that the actions on the pseudovacuum of the new operators $\mu$ and $\nu$ - where new means every time they differ from the original chirality-assigment in section 3.6 - is exactly the same as in (\ref{exac}). For instance, the analogue of $\mu_{12}$ but for $R_{+-} = V^{-1} R_{++} V^{-1}$ is $-i \mbox{sech} \Big(\frac{\gamma_0 - \gamma_{2}}{2}\Big) \, E_{12}$, however the analogue of $\nu_{21}$ but for $R_{--}^{op}=V R^{op}_{+-} V$ is $- \mbox{sech} \Big(\frac{\gamma_0 + \gamma_{2}}{2}\Big) \, E_{21}$, so the action on the pseudovacuum of the product $\mu_{12} \nu_{21}$ is the same as in (\ref{exac}). Similarly is for the other products' actions. This means that the proof in section 3.6 is completely identical for this new arrangement of chiralities, hence $\lambda_2$ is precisely the same. 
Other assignements of chiralities will work out similarly. 

This seems to suggest that a series of accidental identities between the different $R$-matrices is ultimately responsible for what we have observed - it would be very interesting to explore whether there is any more profound physical reason behind this.

}

\section{Conclusions}

In this paper we have {{} studied} the boundary algebraic Bethe ansatz for the massless sector of the $AdS_3 \times S^3 \times T^4$ integrable string. We have {{} constructed the  dual equation} \cite{Sklyanin:1988yz} which ensures the commutativity of the double-row transfer matrix, making it a generating function of the commuting charges in involution. We then applied the method of creating magnons out of the pseudo-vacuum, using the entries of the monodromy - mindful that in the boundary case we have two monodromies $T_+$ and $T_-$ which have to be carefully employed. We have given a step-by-step derivation using the $L$ representation and the singlet boundary, hoping that this serves as a pedagogical account. A series of {{} identities and simplifications, as well as the familiar armamentary of the algebraic Bethe ansatz for six-vertex models, allowed us to obtain} the general form of the Bethe equations, for any number of sites (quantum particles) and excitations (magnons). The final epressions can easily be supplemented with the appropriate strings of dressing factors (omitted here for clarity of notation) if needed. 

Several issues are left open by our analysis:

\begin{itemize}

\item The dual equation is {{} the object of extensive study}\footnote{{} In addition to the literature quoted in the Introduction, there are numerous case-by-case constructions, see for instance \cite{case}. We thank Chiara Paletta for discussions and for pointing out many useful references to us.}, and its construction is done basically with the  purpose to achieve commutativity of the transfer matrix, which is exactly what we did. In bosonic models, as Sklyanin originally found, the dual equation is truly connected with a crossing transformation of the $R$-matrices which appear. We made our proof of commutativity without explicitly referring to the crossing properties of the $AdS_3$ $R$-matrix, and in the process we introduced a number of triple super-transpositions which allow us to invert some relations. 
{{} We} have attempted to reduce/recast our dual equation to a form where crossing symmetry might be manifest, but we have not yet managed to do that. There is the logical possibility that they are not related, but we believe that perhaps a more complicated new rewriting might in the future show this link.     
\item In our massless situation, we discovered in \cite{Bielli:2024xuv} that not only right and left bulk movers have to be treated independently and kept track of very carefully, but also that there is an entire analogous boundary scattering theory which can be defined using a (genuinely) left wall, onto which left movers crash and bounce. We completely constructed such left boundary scattering theory, but never figured out whether there is any relation to the right wall. In the light of the results of this paper, we are led to wonder what the relation might be, if any, between the (literal) left theory of \cite{Bielli:2024xuv}, and the dual equation which we have here.  We also believe that the spectrum one would obtain by applying the algebraic Bethe ansatz to the (genuinely) left wall of \cite{Bielli:2024xuv} would be the same as what we get here, with the same eigenvectors and with the commuting charges just being reshuffled.

\item The eigenvalue $\lambda_2$ appears to have a universal formula which is the same for any assortment of right or left mover quantum particles, in such a way that its expression, and consequently the Bethe equations, can be compactly written in one line by simply analytically extending all the quantum variables to the entire complex plane. We have not found a {{} physical} reason for this occurrence, and only a mathematical reason based on accidental identities between the $R$-matrices. As always it would be very interesting to unravel {{} a physical} reason, possibly along the lines of \cite{Nepomechie:2008ab} - {{}perhaps the accidental identities will turn out to not be accidental after all}.

\item Having demonstrated the viability of the procedure, it will be necessary in future work to systematically exhaust the entire representation theory, in particular to tackle the vector boundary which at the moment appears technically daunting.

\item The Bethe ansatz opens up the way to study the Thermodynamic Bethe ansatz, and, since this is a massless theory, we expect that this might open a window on an associated boundary CFT, presumably in some limit akin to the BMN regime \cite{DiegoBogdanAle}. We intend to explore these exciting avenues in future work.

\end{itemize}

\section{Acknowledgments}

We are indebted with Chiara Paletta for many illuminating discussions on the dual boundary equation, for carefully reading the manuscript and for providing invaluable suggestions. {{} We are thankful to Robert Weston for very interesting discussions and email exchange, originated at the conference {\it In Tropea 2024}. We are extremely grateful to the anonymous referee, who produced extensive notes with suggestions and recommendations.} The work of AT has been
supported by EPSRC-SFI under the (expired) grant EP/S020888/1. 
VM is supported by the STFC under the grant ST/X508809/1. DB has been supported by Thailand NSRF via PMU-B, grant number B13F670063.

\section*{Data and Licence Management}
No additional research data beyond the data presented and cited in this work are needed to validate the research findings in this work. For the purpose of open access, the authors have applied a Creative Commons Attribution (CC BY) licence to any Author Accepted Manuscript version arising.

\begin{appendix}

\section{The dual reflection equation}\label{appendo}
\subsection{Preliminaries}

We will focus here on particles in the $L$ representation.

\paragraph{General Identities}
We denote the two-dimensional unit matrices $E_{ij}$ as
\begin{eqnarray*}
E_{11}=b_1, \qquad E_{12}=f_1, \\ 
E_{22}=b_2, \qquad E_{21}=f_2,
\end{eqnarray*}
to clearly distinguish the bosonic from the fermionic ones. The relations that follow hold for generic matrices of the form
\begin{eqnarray}
&R=\beta_{ij}\,b_i\otimes b_j+\varphi_{ij}\, f_i\otimes f_j,\\
&T=B_i\otimes b_i+F_i\otimes f_i,
\end{eqnarray}
where $\beta_{ij}$ and $\varphi_{ij}$ are arbitrary bosonic coefficients and $B_i$, $F_i$ are arbitrary bosonic and fermionic matrices. The symbol $(st)^3$ denotes triple supertransposition, i.e. supertransposing the same space three times. The identities used are
\begin{align}
\text{str}_1(T_1^+T_1^-)&=\text{str}_1((T_1^+)^{st_1}(T_1^-)^{st_1})\label{id1},
\\
R_{12}^{st_1}T_1^{st_1}T_2&=(T_1R_{12}T_2)^{st_1}\label{id2},
\\
T_1T_2R_{12}^{st_2}&=(T_1R_{12}T_2^{(st_2)^3})^{st_2},\label{id3}
\\
\text{str}_{12}((T^+_1R_{12}T^+_2)^{st_2}(T^-_1\tilde{R}_{12}T^-_2)^{st_1})&=\text{str}_{12}(((T^+_1R_{12}T^+_2)^{st_2}(T^-_1\tilde{R}_{12}T^-_2)^{st_1})^{(st_1)^3})
\nonumber\\
&=\text{str}_{12}(((T^+_1R_{12}T^+_2)^{st_2})^{(st_1)^3}T^-_1\tilde{R}_{12}T^-_2),\label{id4}
\\
((T_1R_{12}T_2)^{st_2})^{(st_1)^3}(\tilde{R}^{st_2}_{12})^{(st_1)^3}&=((\tilde{R}_{12}T_1 R_{12}T_2)^{st_2})^{(st_1)^3},
\label{id5} \\
\text{str}_{12}(\tilde{R}_{12}T_2^+T_1^+T_2^-T_1^-R_{12})&=\text{str}_{12}(T_2^+T_1^+T_2^-T_1^-R_{12}\tilde{R}_{12}),
\label{id6} \\
\text{str}_{12}(R_{12}((T_2^+\tilde{R}_{12}T_1^+)^{st_2})^{(st_1)^3}T_2^-R'_{12}T_1^-\tilde{R}'_{12})&=\text{str}_{12}(((T_2^+\tilde{R}_{12}T_1^+)^{st_2})^{(st_1)^3}T_2^-R'_{12}T_1^-\tilde{R}'_{12}R_{12}),
\label{id7} \\
\text{str}_{12}(((T_1^+R_{12}T_2^-)^{st_2})^{(st_1)^3}T_1^-\tilde{R}_{12}T_2^-)&=\text{str}_{12}((((T_1^+R_{12}T_2^-)^{st_2})^{(st_1)^3}T_1^-\tilde{R}_{12}T_2^-)^{st_1})\nonumber
\\
&=\text{str}_{12}((T_1^+R_{12}T_2^-)^{st_2}(T_1^-\tilde{R}_{12}T_2^-)^{st_1}),
\label{id8}
\end{align}
all of which we have checked with the help of Mathematica, keeping track of the fermionic signs and the graded tensor products.
\paragraph{R-matrix relations}
In the following relations, $R_{xy}$, $x,y = \pm$, denotes one of the $L-L$ $R$-matrices and $\Sigma_2=\mathbf{1}\otimes \sigma_3$, where $\sigma_3$ is the diagonal Pauli matrix.
\begin{align}
&R_{+-}^{st_2}(\gamma)R_{+-}^{st_1}(\gamma)=\text{tanh}^2\Big(\frac{\gamma}{2}\Big) \, \mathbbmss{1} \otimes \mathbbmss{1},\label{rel1}\\
&((\Sigma_2 R_{--}^{op}(\gamma)\Sigma_2)^{st_2})^{(st_1)^3}R_{--}^{op}(\gamma)=\mathbbmss{1} \otimes \mathbbmss{1},\label{rel2}\\
&R_{++}(\gamma)((\Sigma_2 R_{++}(\gamma)\Sigma_2)^{st_2})^{(st_1)^3}=\mathbbmss{1} \otimes \mathbbmss{1},\label{rel3}\\
&(R^{op}_{+-}(\gamma))^{st_2}(R_{+-}^{op}(\gamma))^{st_1}=\text{tanh}^2\Big(\frac{\gamma}{2}\Big)\, \mathbbmss{1} \otimes \mathbbmss{1}.\label{rel4}
\end{align}
\paragraph{The right-wall reflection equation}
To simplify notation, we rewrite the singlet-boundary right-wall reflection equation \eqref{BYBE_original_form} as
\begin{align}
T^-_2(\gamma_2)R^{op}_{+-}(\gamma_+)T^-_1(\gamma_1)R_{++}(\gamma_-)=R^{op}_{--}(\gamma_-)T^-_1(\gamma_1)R_{+-}(\gamma_+)T_2^-(\gamma_2),\label{BYBEa}
\end{align}
where $\gamma_\pm=\gamma_1\pm\gamma_2$.
\paragraph{The dual reflection equation}
The dual BYBE, which ensures the commutativity of the transfer matrix, will be proven in the next subsection to be
\begin{align}
\Sigma_2R^{op}_{--}(\gamma_-)\Sigma_2(T^+_1(\gamma_1))^{st_1}&R_{+-}(\gamma_+)(T_2^+(\gamma_2))^{(st_2)^3}=\nonumber\\
&(T_2^+(\gamma_2))^{(st_2)^3}R^{op}_{+-}(\gamma_+)(T_1^+(\gamma_1))^{st_1}\Sigma_2R_{++}(\gamma_-)\Sigma_2.\label{dual_BYBE}
\end{align} 

\paragraph{Comparison with left-wall BYBE}
This dual reflection equation can be brought to a slightly different form by noticing that 
\begin{align*}
\Sigma_2 R^{op}_{--}(\gamma)\Sigma_2=R^{op}_{--}(\gamma-2 i \pi),\\
\Sigma_2 R_{++}(\gamma)\Sigma_2=R_{++}(\gamma-2 i \pi),
\end{align*} 
and transposing both sides \eqref{dual_BYBE} as $(\cdot)^{(st_1)^3st_2}$, leading to \begin{align*}
T^+_2(\gamma_2)(R_{+-}(\gamma_+))^{(st_1)^3st_2}T^+_1(\gamma_1)&(R_{--}^{op}(\gamma_--2 i\pi ))^{(st_1)^3st_2}=\\&(R_{++}(\gamma_--2 i\pi ))^{(st_1)^3st_2}T^+_1(\gamma_1)(R^{op}_{+-}(\gamma_+))^{(st_1)^3st_2}T^+_2(\gamma_2)
\end{align*}
(which we have checked with Mathematica).
Using some further properties of the $L-L$ $R$-matrix, this is equivalent to 
\begin{align}
T^+_2(\gamma_2)R_{+-}(\gamma_+\pm 2 i \pi)T^+_1(\gamma_1)R_{--}^{op}(\gamma_- )=R_{++}(\gamma_-)T^+_1(\gamma_1)R^{op}_{+-}(\gamma_+\pm 2 i \pi)T^+_2(\gamma_2).
\end{align}
As a comparison, the left-wall BYBE in \cite{Bielli:2024xuv} is
\begin{align*}
T_2(-\tilde{\gamma}_2)R_{+-}(\tilde{\gamma}_+)T_1(-\tilde{\gamma}_1)R_{--}^{op}(\tilde{\gamma}_-)=R_{++}(\tilde{\gamma}_-)T_1(-\tilde{\gamma}_1)R_{+-}^{op}(\tilde{\gamma}_+)T_2(-\tilde{\gamma}_2),
\end{align*}
where $\tilde{\gamma}_i>0$. Rewriting it using $\gamma_i=-\tilde{\gamma}_i<0$ we have
\begin{align}\label{left_wall_BYBE}
T_2(\gamma_2)R_{+-}(-\gamma_+)T_1(\gamma_1)R_{--}^{op}(-\gamma_-)=R_{++}(-\gamma_-)T_1(\tilde{\gamma}_1)R_{+-}^{op}(-\gamma_+)T_2(\gamma_2).
\end{align}
One of the differences between \eqref{left_wall_BYBE} and \eqref{dual_BYBE}
is the relative sign between the rapidity arguements in the monodromies and the $R$-matrices: in \eqref{left_wall_BYBE} the rapidities appear with opposite signs in $T$ and $R$ while in \eqref{dual_BYBE} with the same sign. Comparing with Sklyanin's (13) in \cite{Sklyanin:1988yz}, the shift in the $R$-matrix appears as expected but we have the same issue of the sign of rapidities, which cannot be resolved by reparemetrizing $\gamma_i\rightarrow -\gamma_i$.
\subsection{Proof of the commutativity of the transfer matrix}
We define the transfer matrix of our open system using two single-row monodromies $T^+(\gamma_i)$ and $T^-(\gamma_i)$ which solve the dual and original BYBE respectively. To declutter the expressions that follow, we omit the rapidity argument of the monodromies while keeping it for the $R$-matrices. Assuming that the parts of $T^+$ and $T^-$ that act on the physical space commute\footnote{This property will be guaranteed in all the representations of our interest by taking $T_+$ as acting only in the auxiliary space as $K_+$, and acting as identity in the quantum physical spaces.}, the product of two transfer matrices can be written as
\begin{align*}
    t(\gamma_1)t(\gamma_2)&= \text{str}_1(T^+_1T^-_1)\text{str}_2(T^+_2T^-_2)\\
    &=\text{str}_1((T^+_1)^{st_1}(T^-_1)^{st_1})\text{str}_2(T^+_2T^-_2)\\
    &=\text{str}_{12}((T^+_1)^{st_1}(T^-_1)^{st_1}T^+_2T^-_2)\\
    &=\text{str}_{12}((T^+_1)^{st_1}T^+_2(T^-_1)^{st_1}T^-_2),
\end{align*}
where the identity \eqref{id1} has been used. Using the  $R$-matrix relation \eqref{rel1} and then the identities \eqref{id2}, \eqref{id3} followed by \eqref{id4}, we obtain
\begin{align*}
t(\gamma_1)t(\gamma_2)&=\frac{1}{\text{tanh}^2(\frac{\gamma_+}{2})}\,\text{str}_{12}((T^+_1)^{st_1}T^+_2R_{+-}^{st_2}(\gamma_+)R_{+-}^{st_1}(\gamma_+)(T^-_1)^{st_1}T^-_2)\\
&=\frac{1}{\text{tanh}^2(\frac{\gamma_+}{2})}\,\text{str}_{12}(((T^+_1)^{st_1}R_{+-}(\gamma_+)(T^+_2)^{({st_2})^3})^{st_2}(T^-_1R_{+-}(\gamma_+)T^-_2)^{st_1})\\
&=\frac{1}{\text{tanh}^2(\frac{\gamma_+}{2})}\,\text{str}_{12}(((T^+_1)^{st_1}R_{+-}(\gamma_+)(T^+_2)^{({st_2})^3})^{st_2\,(st_1)^3}T^-_1R_{+-}(\gamma_+)T^-_2),
\end{align*}
We now insert a second pair of $R$-matrices using \eqref{rel2} and apply \eqref{id5}:
\begin{align*}
&t(\gamma_1)t(\gamma_2)\\
&=\frac{1}{\text{tanh}^2\!(\frac{\gamma_{\!+\!}}{2})\!}\,\text{str}_{\!12\!}(((T^+_1)^{st_1}R_{+-}(\gamma_{\!+\!})(T^+_2)^{({st_2})^3})^{st_2\,(st_1)^3}(\Sigma_2 R_{--}^{op}(\gamma_{\!-\!})\Sigma_2^{st_2\,(st_1)^3})R_{--}^{op}(\gamma_{\!-\!})T^-_1R_{+-}(\gamma_{\!+\!})T^-_2)\\
&=\frac{1}{\text{tanh}^2\!(\frac{\gamma_{\!+\!}}{2})\!}\,\text{str}_{\!12\!}((\Sigma_2 R_{--}^{op}(\gamma_{\!-\!})\Sigma_2(T^+_1)^{st_1}R_{+-}(\gamma_{\!+\!})(T^+_2)^{({st_2})^3})^{st_2\,(st_1)^3}R_{--}^{op}(\gamma_{\!-\!})T^-_1R_{+-}(\gamma_{\!+\!})T^-_2),
\end{align*}
What is inside the trace in the last expression is of the form $(A^{st_2\,st_1^3}B)$, where $A$ is one side of the dual BYBE \eqref{dual_BYBE} and $B$ is one side of the original BYBE \eqref{BYBE}. We can now apply both of these relations followed by \eqref{id5}, \eqref{id7} and \eqref{id6}, in that order:
\begin{align*}
&t(\gamma_1)t(\gamma_2)\\
& \!=\! \frac{1}{\text{tanh}^2\!(\frac{\gamma_{\!+\!}}{2})\!}\,\text{str}_{\!12\!}(((T^+_2)^{(st_2)^3}R_{+-}^{op}(\gamma_{\!+\!})(T_1^+)^{st_1}\Sigma_2 R_{++}(\gamma_{\!-\!})\Sigma_2)^{st_2\,(st_1)^3}T^-_2R^{op}_{+-}(\gamma_{\!+\!})T^-_1R_{++}(\gamma_{\!-\!}))\\
& \!=\! \frac{1}{\text{tanh}^2\!(\frac{\gamma_{\!+\!}}{2})\!}\,\text{str}_{\!12\!}((\Sigma_2 R_{++}(\gamma_{\!-\!})\Sigma_2)^{st_2\,(st_1)^3}((T^+_2)^{(st_2)^3}R_{+-}^{op}(\gamma_{\!+\!})(T_1^+)^{st_1})^{st_2\,(st_1)^3}T^-_2R^{op}_{+-}(\gamma_{\!+\!})T^-_1R_{++}(\gamma_{\!-\!}))\\
& \!=\! \frac{1}{\text{tanh}^2\!(\frac{\gamma_{\!+\!}}{2})\!}\,\text{str}_{\!12\!}(((T^+_2)^{(st_2)^3}R_{+-}^{op}(\gamma_{\!+\!})(T_1^+)^{st_1})^{st_2\,(st_1)^3}T^-_2R^{op}_{+-}(\gamma_{\!+\!})T^-_1R_{++}(\gamma_{\!-\!})(\Sigma_2 R_{++}(\gamma_{\!-\!})\Sigma_2)^{st_2\,(st_1)^3}))\\
& \!=\! \frac{1}{\text{tanh}^2\!(\frac{\gamma_{\!+\!}}{2})\!}\,\text{str}_{\!12\!}(((T^+_2)^{(st_2)^3}R_{+-}^{op}(\gamma_{\!+\!})(T_1^+)^{st_1})^{st_2\,(st_1)^3}T^-_2R^{op}_{+-}(\gamma_{\!+\!})T^-_1),
\end{align*}
where \eqref{rel3} was used in the last equality. Using the remaining identity \eqref{id8} and then reusing \eqref{id2}, \eqref{id3} we have
\begin{align*}
t(\gamma_1)t(\gamma_2)&=\frac{1}{\text{tanh}^2(\frac{\gamma_+}{2})}\,\text{str}_{12}(((T^+_2)^{(st_2)^3}R_{+-}^{op}(\gamma_+)(T_1^+)^{st_1})^{st_2}(T^-_2R^{op}_{+-}(\gamma_+)T^-_1)^{st_1})\\
&=\frac{1}{\text{tanh}^2(\frac{\gamma_+}{2})}\,\text{str}_{12}((R_{+-}^{op}(\gamma_+))^{st_2}T^+_2(T_1^+)^{st_1}T^-_2(T^-_1)^{st_1}(R^{op}_{+-}(\gamma_+))^{st_1}).
\end{align*}
Finally, we can use \eqref{id6} and \eqref{rel4}: 
\begin{align*}
t(\gamma_1)t(\gamma_2)&=\frac{1}{\text{tanh}^2(\frac{\gamma_+}{2})}\,\text{str}_{12}(T^+_2(T_1^+)^{st_1}T^-_2(T^-_1)^{st_1}(R^{op}_{+-}(\gamma_+))^{st_1}(R_{+-}^{op}(\gamma_+))^{st_2})\\
&=\text{str}_{12}(T^+_2(T_1^+)^{st_1}T^-_2(T^-_1)^{st_1})\\
&=\text{str}_{12}(T^+_2T^-_2(T_1^+)^{st_1}(T^-_1)^{st_1})\\
&=t(\gamma_2)t(\gamma_1),
\end{align*}
where we again assume that $T^-_2$ and $(T_1^+)^{st_1}$ commute, as in the very beginning of this calculation.

\section{Proof that $\lambda_1(\gamma)=1$ for any $N$}\label{B}
This can be easily achieved by writing $\lambda_1(\gamma_0)$ as
\begin{eqnarray}
\lambda_1(\gamma_0)|0\rangle =  
\langle \phi | T_-(\gamma_0)|\phi\rangle \otimes \big[ |\phi\rangle \otimes |\phi\rangle... \otimes |\phi\rangle],    
\end{eqnarray}
where the ket and the first bra vectors are in the space $0$, and the remaining ket vectors are the $N$ physical spaces. Now we recall that $T_-$ is given by (\ref{Tminus}), which we reproduce here for ease of reading:
\begin{eqnarray}
&&T_- = [R_{++}]_{0,N}(\gamma_0-\gamma_N)\,[R_{++}]_{0,N-1}(\gamma_0-\gamma_{N-1})...[R_{++}]_{0,2}(\gamma_0-\gamma_2)\,[R_{++}]_{0,1}(\gamma_0-\gamma_1)K(\gamma_0)_0 \nonumber\\
&&\qquad \qquad \times[R^{op}_{+-}]_{0,1}(\gamma_1+\gamma_0)\,[R^{op}_{+-}]_{0,2}(\gamma_2+\gamma_0)...[R^{op}_{+-}]_{0,N-1}(\gamma_{N-1}+\gamma_0)\,[R^{op}_{+-}]_{0,N}(\gamma_N+\gamma_0).\nonumber  
\end{eqnarray}
It is easy to see that, in our conventions and ignoring the dressing factors, all the $R_{xy}$-matrices, $x,y = \pm$ (no matter which assortment of signs) act on the state $|\phi\rangle \otimes |\phi\rangle$ as $1 \times |\phi\rangle \otimes |\phi\rangle$. This means that the string of $R$-matrices leaves the ket unchanged. Moreover, the reflection matrix is diagonal, and also has a $1$ in the boson-boson entry. This means that the ket goes unchanged through $T_-$, and the bra makes $\langle \phi|\phi\rangle = 1$ in the auxiliary $0$ space, leaving the pseudovacuum multiplied by $1$ in the physical space. 

{{}
\section{Proof of (\ref{BYBE2})\label{BYBE2section}}

In this appendix we prove the $N=1$ case of (\ref{BYBE2}), the general-$N$ case following by induction.

We start by explicitly writing the l.h.s. of \eqref{BYBE2} using the definition \eqref{Tminus} for $N=1$:
\begin{equation}\label{step-1}
\begin{aligned}
&T^{0'}_-(\gamma_{0'}) [R^{op}_{+-}]_{0,0'}\Big(\gamma_{0'} - (-\gamma_0)\Big)T^0_-(\gamma_0)[R_{++}]_{0,0'}\Big( \gamma_0-\gamma_{0'}\Big) =
\\
& = [R_{++}]_{0',1}(\gamma_{0'}-\gamma_1) \, K_{0'}(\gamma_{0'}) \, [R^{op}_{+-}]_{0',1}(\gamma_{0'}+\gamma_1) \, [R^{op}_{+-}]_{0,0'}(\gamma_{0'}+\gamma_0) \,[R_{++}]_{0,1}(\gamma_{0}-\gamma_1) \times \\ 
&\qquad \qquad \qquad \qquad\qquad \qquad\qquad \qquad\quad \times K_0(\gamma_0) \, [R^{op}_{+-}]_{0,1}(\gamma_{0}+\gamma_1) \, [R_{++}]_{0,0'}(\gamma_{0}-\gamma_{0'}) \,\, .
\end{aligned}
\end{equation}
We now need to use two relations for our $R$-matrices, which can be checked by brute force
\begin{eqnarray}\label{relan1}
R^{op}_{+-}(\gamma) = R^{-1}_{-+}(-\gamma),
\qquad \qquad 
R^{op}_{+-}(\gamma) = R_{+-}(\gamma) \,\, .
\end{eqnarray}
Thanks to these we can rewrite \eqref{step-1} as
\begin{equation}\label{step-2}
\begin{aligned}
&[R_{++}]_{0',1}(\gamma_{0'}-\gamma_1) \, K_{0'}(\gamma_{0'}) \, [R^{-1}_{-+}]_{0',1}(-\gamma_{0'}-\gamma_1) \, [R_{+-}]_{0,0'}(\gamma_{0}-(-\gamma_{0'})) \,[R_{++}]_{0,1}(\gamma_{0}-\gamma_1) \times 
\\ 
&\qquad \qquad \qquad \qquad\qquad \qquad\qquad \qquad\quad \times K_0(\gamma_0) \, [R^{op}_{+-}]_{0,1}(\gamma_{0}+\gamma_1) \, [R_{++}]_{0,0'}(\gamma_{0}-\gamma_{0'}) \,\, .
\end{aligned}
\end{equation}
It is now possible to use the YBE in the form $R_{23}^{-1}(\gamma_2-\gamma_3) R_{12}(\gamma_1-\gamma_2) R_{13}(\gamma_1-\gamma_3) = R_{13}(\gamma_1-\gamma_3) R_{12}(\gamma_1-\gamma_2)R^{-1}_{23}(\gamma_2-\gamma_3)$, with the roles of spaces $1-2-3$ played by $0-0'-1$ and the role of variables $\gamma_1,\gamma_2,\gamma_3$ played by $\gamma_0,-\gamma_{0'},\gamma_1$ respectively:
\begin{equation}
\begin{aligned}
&[R^{-1}_{-+}]_{0',1}(-\gamma_{0'}-\gamma_1) \, [R_{+-}]_{0,0'}(\gamma_{0}-(-\gamma_{0'})) \,[R_{++}]_{0,1}(\gamma_{0}-\gamma_1) = 
\\
&  \qquad \qquad = [R_{++}]_{0,1}(\gamma_{0}-\gamma_1) \, [R_{+-}]_{0,0'}(\gamma_{0}-(-\gamma_{0'})) \,[R^{-1}_{-+}]_{0',1}(-\gamma_{0'}-\gamma_1) \,\, .  
\end{aligned}
\end{equation}
We therefore get, from \eqref{step-2}
\begin{equation}
\begin{aligned}
&[R_{++}]_{0',1}(\gamma_{0'}-\gamma_1) \, K_{0'}(\gamma_{0'}) \, [R_{++}]_{0,1}(\gamma_{0}-\gamma_1) \, [R_{+-}]_{0,0'}(\gamma_{0}-(-\gamma_{0'})) \,[R^{-1}_{-+}]_{0',1}(-\gamma_{0'}-\gamma_1) \times
\\ 
& \qquad \qquad \qquad \qquad\qquad \qquad\qquad \qquad\quad \times K_0(\gamma_0) \, [R^{op}_{+-}]_{0,1}(\gamma_{0}+\gamma_1) \, [R_{++}]_{0,0'}(\gamma_{0}-\gamma_{0'}) \,\, ,
\end{aligned}
\end{equation}
which, by using again (\ref{relan1}), can be converted back to
\begin{equation}
\begin{aligned}
&[R_{++}]_{0',1}(\gamma_{0'}-\gamma_1) \, K_{0'}(\gamma_{0'}) \, [R_{++}]_{0,1}(\gamma_{0}-\gamma_1) \, [R^{op}_{+-}]_{0,0'}(\gamma_{0}-(-\gamma_{0'})) \,[R^{op}_{+-}]_{0',1}(\gamma_{0'}+\gamma_1) \times 
\\
&\qquad \qquad \qquad \qquad\qquad \qquad\qquad \qquad\quad \times K_0(\gamma_0) \, [R^{op}_{+-}]_{0,1}(\gamma_{0}+\gamma_1) \, [R_{++}]_{0,0'}(\gamma_{0}-\gamma_{0'}) \,\, .
\end{aligned}
\end{equation}
Having spelt out one step where the YBE is used, we will now be more concise on the remaining steps. We commute $K_0$ and $[R^{op}_{+-}]_{0',1}$ with no sign issue since $K_0$ is bosonic in the singlet case, obtaining 
\begin{equation}
\begin{aligned}
&[R_{++}]_{0',1}(\gamma_{0'}-\gamma_1) \, K_{0'}(\gamma_{0'}) \, [R_{++}]_{0,1}(\gamma_{0}-\gamma_1) \, [R^{op}_{+-}]_{0,0'}(\gamma_{0}-(-\gamma_{0'})) \, K_0(\gamma_0) \, \times 
\\ 
& \qquad \qquad \qquad \qquad\qquad  \times [R^{op}_{+-}]_{0',1}(\gamma_{0'}+\gamma_1)\, [R^{op}_{+-}]_{0,1}(\gamma_{0}+\gamma_1)  \, [R_{++}]_{0,0'}(\gamma_{0}-\gamma_{0'}) \,\, .
\end{aligned}
\end{equation}
We now use again that $R^{op}_{+-}(\gamma) = R_{+-}(\gamma)$ and then the YBE 
\begin{equation}
\begin{aligned}
& [R_{+-}]_{0',1}(\gamma_{0'}+\gamma_1)\, [R_{+-}]_{0,1}(\gamma_{0}+\gamma_1)  \, [R_{++}]_{0,0'}(\gamma_{0}-\gamma_{0'}) =  
\\
&\qquad \qquad  = [R_{++}]_{0,0'}(\gamma_{0}-\gamma_{0'})\,[R_{+-}]_{0,1}(\gamma_{0}+\gamma_1)\,[R_{+-}]_{0',1}(\gamma_{0'}+\gamma_1) \,\, ,
\end{aligned}
\end{equation}
to obtain
\begin{equation}
\begin{aligned}
&[R_{++}]_{0',1}(\gamma_{0'}-\gamma_1) \, K_{0'}(\gamma_{0'}) \, [R_{++}]_{0,1}(\gamma_{0}-\gamma_1) \, [R^{op}_{+-}]_{0,0'}(\gamma_{0}-(-\gamma_{0'})) \, K_0(\gamma_0) \, \times
\\
& \qquad \qquad \qquad \qquad\qquad  \times [R_{++}]_{0,0'}(\gamma_{0}-\gamma_{0'})\,[R^{op}_{+-}]_{0,1}(\gamma_{0}+\gamma_1)\,[R^{op}_{+-}]_{0',1}(\gamma_{0'}+\gamma_1) \,\, ,
\end{aligned}
\end{equation}
and then commute $K_{0'}$, arriving to
\begin{equation}
\begin{aligned}
&[R_{++}]_{0',1}(\gamma_{0'}-\gamma_1) \,  [R_{++}]_{0,1}(\gamma_{0}-\gamma_1) \,K_{0'}(\gamma_{0'}) \, [R^{op}_{+-}]_{0,0'}(\gamma_{0}-(-\gamma_{0'})) \, K_0(\gamma_0) \, \times
\\ 
&\qquad \qquad \qquad \qquad\qquad  \times [R_{++}]_{0,0'}(\gamma_{0}-\gamma_{0'})\,[R^{op}_{+-}]_{0,1}(\gamma_{0}+\gamma_1)\,[R^{op}_{+-}]_{0',1}(\gamma_{0'}+\gamma_1) \,\, .
\end{aligned}
\end{equation}
Now we use the BYBE (\ref{BYBE}) (in the spaces $0-0'-1$) and turn the above into 
\begin{align}\label{step-3}
&[R_{++}]_{0',1}(\gamma_{0'}-\gamma_1) \,  [R_{++}]_{0,1}(\gamma_{0}-\gamma_1) \,[R^{op}_{--}]_{0,0'}\Big((-\gamma_{0'})-(-\gamma_0)\Big)K_0(\gamma_0)[R_{+-}]_{0,0'}\Big(\gamma_0 - (-\gamma_{0'})\Big) \times
\notag \\ 
&\qquad \qquad \qquad \qquad\qquad  \times \,K_{0'}(\gamma_{0'}) \,[R^{op}_{+-}]_{0,1}(\gamma_{0}+\gamma_1)\,[R^{op}_{+-}]_{0',1}(\gamma_{0'}+\gamma_1) \,\, .
\end{align}
We further use the property (which can again be verified by brute force)
\begin{eqnarray}
R^{op}_{--}(\gamma) = R_{++}(\gamma) \,\, ,
\end{eqnarray}
and then the YBE
\begin{equation}
\begin{aligned}
& [R_{++}]_{0',1}(\gamma_{0'}-\gamma_1) \,  [R_{++}]_{0,1}(\gamma_{0}-\gamma_1) \,[R_{++}]_{0,0'}(\gamma_0 - \gamma_{0'}) = 
\\
&\qquad \qquad = [R_{++}]_{0,0'}(\gamma_0 - \gamma_{0'}) \, [R_{++}]_{0,1}(\gamma_{0}-\gamma_1)\,[R_{++}]_{0',1}(\gamma_{0'}-\gamma_1) \,\, ,
\end{aligned}
\end{equation}
to turn \eqref{step-3} into
\begin{align}
& [R^{op}_{--}]_{0,0'}\Big((-\gamma_{0'})-(-\gamma_0)\Big)\,  [R_{++}]_{0,1}(\gamma_{0}-\gamma_1) \,[R_{++}]_{0',1}(\gamma_{0'}-\gamma_1)K_0(\gamma_0)[R_{+-}]_{0,0'}\Big(\gamma_0 - (-\gamma_{0'})\Big) \times 
\notag \\ 
&\qquad \qquad \qquad \qquad\qquad  \times \,K_{0'}(\gamma_{0'}) \,[R^{op}_{+-}]_{0,1}(\gamma_{0}+\gamma_1)\,[R^{op}_{+-}]_{0',1}(\gamma_{0'}+\gamma_1) \,\, .
\end{align}
We can now commute both $K$-matrices:
\begin{align}
& [R^{op}_{--}]_{0,0'}\Big((-\gamma_{0'})-(-\gamma_0)\Big)\,  [R_{++}]_{0,1}(\gamma_{0}-\gamma_1) \,K_0(\gamma_0)[R_{++}]_{0',1}(\gamma_{0'}-\gamma_1)[R_{+-}]_{0,0'}\Big(\gamma_0 - (-\gamma_{0'})\Big) \times 
\notag  \\
&\qquad \qquad \qquad \qquad\qquad  \times \, \,[R^{op}_{+-}]_{0,1}(\gamma_{0}+\gamma_1)K_{0'}(\gamma_{0'})\,[R^{op}_{+-}]_{0',1}(\gamma_{0'}+\gamma_1).
\end{align}
Finally we use $[R_{++}]_{0',1}(\gamma_{0'} - \gamma_1) = [R^{-1}_{--}]_{0',1}(\gamma_1 - \gamma_{0'})$ (once again checked by brute force) and the YBE in the form
$R_{23}^{-1}(\gamma_2-\gamma_3) R_{12}(\gamma_1-\gamma_2) R_{13}(\gamma_1-\gamma_3) = R_{13}(\gamma_1-\gamma_3) R_{12}(\gamma_1-\gamma_2)R^{-1}_{23}(\gamma_2-\gamma_3)$, with the roles of spaces $1-2-3$ played by $0-0'-1$ and the role of variables $\gamma_1,\gamma_2,\gamma_3$ played by $\gamma_0,-\gamma_{0'},-\gamma_1$ respectively, to get
\begin{equation}
\begin{aligned}
& [R^{op}_{--}]_{0,0'}\Big((-\gamma_{0'})-(-\gamma_0)\Big)\,  [R_{++}]_{0,1}(\gamma_{0}-\gamma_1) \,K_0(\gamma_0)[R^{op}_{+-}]_{0,1}(\gamma_{0}+\gamma_1)[R_{+-}]_{0,0'}\Big(\gamma_0 - (-\gamma_{0'})\Big) \times 
\\
&\qquad \qquad \qquad \qquad\qquad  \times \, [R_{++}]_{0',1}(\gamma_{0'}-\gamma_1)\,K_{0'}(\gamma_{0'})\,[R^{op}_{+-}]_{0',1}(\gamma_{0'}+\gamma_1) = 
\\
& = [R^{op}_{--}]_{0,0'}\Big((-\gamma_{0'})-(-\gamma_0)\Big)T^0_-(\gamma_0)[R_{+-}]_{0,0'}\Big(\gamma_0 - (-\gamma_{0'})\Big)T_{-}^{0'}(\gamma_{0'}) \,\, .
\end{aligned}
\end{equation}
which completes the proof of (\ref{BYBE2}) for $N=1$.

\section{\label{gene}Proof of (\ref{generale}) and of the Bethe equations (\ref{Bethe2})}

We first prove the eigenvalue formula \eqref{generale} by induction. Consider a $(\!M+1\!)$-magnon state and compute
\begin{eqnarray}
\Big[\mbox{str}_0 T_- (\gamma_0)\Big] \prod_{j=1}^{M+1}B(\mu_j)|0\rangle =    \Big[\mbox{str}_0 T_- (\gamma_0)\Big]B(\mu_{M+1})\prod_{j=1}^{M}B(\mu_j)|0\rangle \,\, ,
\end{eqnarray}
where we have used the commutativity of the $B$ operators. It is now easy to use (\ref{rew}) to get
\begin{equation}
\begin{aligned}
& \Big[\mbox{str}_0 T_- (\gamma_0)\Big]B(\mu_{M+1})\prod_{j=1}^{M}B(\mu_j)|0\rangle =  \text{unwanted} 
 \, + 
\\
&  + \frac{1}{2} \mbox{sech} \Big(\frac{\gamma_0+\mu_{M+1}}{2}\Big)\mbox{csch} \Big(\frac{\gamma_0-\mu_{M+1}}{2}\Big)(\sinh \gamma_0 + \sinh \mu_{M+1}) B(\mu_{M+1})\Lambda_M \prod_{j=1}^{M}B(\mu_j)|0\rangle \,\, ,
\end{aligned}
\end{equation}
having used the induction hypothesis that the transfer matrix is diagonal on the $M$-magnon state. The product formula \eqref{generale} for the eigenvalue then follows promptly by the initial condition (pseudovacuum eigenvalue) $\Lambda_0 = 1 - g(\gamma_0)\lambda_2(\gamma_0)$. The unwanted term can be dealt with in a familiar way.

Let us now turn our attention to the proof of \eqref{Bethe2}. To this aim consider 
\begin{eqnarray}\label{for-Bethe-eqs}
\Big[\mbox{str}_0 T_- (\gamma_0)\Big]\prod_{j=1}^{M}B(\mu_j)|0\rangle = \Big[\mbox{str}_0 T_- (\gamma_0)\Big]B(\mu_1) \prod_{j=2}^{M}B(\mu_j)|0\rangle \,\, .
\end{eqnarray}
What we can do is to keep track, in the repeated application of the RTRT relations to move the transfer matrix towards the pseudovacuum, of the specific term which does not involve $B(\mu_1)$. This can only occur if $B(\mu_1)$ changes into $B(\gamma_0)$ at the first iteration, and then each remaining $B(\mu_j)$, $j=2,...,M$, do {\it not} change their argument. Therefore, the total coefficient accrued by virtue of (\ref{use}) is
\begin{equation}\label{expra}
\begin{aligned}
&  \frac{1}{2} {\rm sech}\big(\frac{\gamma_0 + \mu_1}{2}\big) \Big(2 i  \delta_2 (\mu_1) + {\rm csch}\big(\frac{\gamma_0 - \mu_1}{2}\big)\big[   -2 \delta_1(\mu_1) \,{\rm sinh}\big(\frac{\gamma_0 + \mu_1}{2}\big) \big]\Big)+
\\
& - g(\gamma_0) \Bigg(\frac{1}{2} {\rm sech}\big(\frac{\gamma_0 + \mu_1}{2}\big) \Big(-2 i  \delta_1(\mu_1) + {\rm csch}\big(\frac{\gamma_0 - \mu_1}{2}\big)\big[-2 \delta_2(\mu_1) \,{\rm sinh}\big(\frac{\gamma_0 + \mu_1}{2}\big) \big]\Big)\Bigg) \,\, ,
\end{aligned}
\end{equation}
where $\delta_1(\mu_1)$ is only the {\it diagonal} part of the action of $A(\mu_1)$ on $\prod_{j=2}^M B(\mu_j)|0\rangle$, and $\delta_2(\mu_1)$ is the {\it diagonal} part of the action of $D(\mu_1)$ on $\prod_{j=2}^M B(\mu_j)|0\rangle$. These are easily obtained as
\begin{equation}
\begin{aligned}
\delta_1(\mu_1) &= \lambda_{1}(\mu_{1})\prod_{j=2}^M \mbox{csch} \big(\frac{\mu_1 - \mu_j}{2}\big) \, (\sinh \mu_1 + \sinh \mu_j) \,\, ,
\\
\delta_2(\mu_1) &= \lambda_2(\mu_1) \prod_{j=2}^M \mbox{csch} \big(\frac{\mu_1 - \mu_j}{2}\big) \, (\sinh \mu_1 + \sinh \mu_j) \,\, .
\end{aligned}
\end{equation}
This means that the expression (\ref{expra}) has an overall $\prod_{j=2}^M \mbox{csch} \big(\frac{\mu_1 - \mu_j}{2}\big) \, (\sinh \mu_1 + \sinh \mu_j)$, which is immaterial if we are to set this term to zero, and so we are left with
\begin{equation}\label{expras}
\begin{aligned}
&  \frac{1}{2} {\rm sech}\big(\frac{\gamma_0 + \mu_1}{2}\big) \Big(2 i  \lambda_2 (\mu_1) + {\rm csch}\big(\frac{\gamma_0 - \mu_1}{2}\big)\big[   -2 \lambda_1(\mu_1) \,{\rm sinh}\big(\frac{\gamma_0 + \mu_1}{2}\big) \big]\Big) + 
\\
& - g(\gamma_0) \Bigg(\frac{1}{2} {\rm sech}\big(\frac{\gamma_0 + \mu_1}{2}\big) \Big(-2 i  \lambda_1(\mu_1) + {\rm csch}\big(\frac{\gamma_0 - \mu_1}{2}\big)\big[-2 \lambda_2(\mu_1) \,{\rm sinh}\big(\frac{\gamma_0 + \mu_1}{2}\big) \big]\Big)\Bigg) \,\, .
\end{aligned}
\end{equation}
Setting this to zero produces the Bethe equations \eqref{Bethe} in the main text of the paper for $\mu_1$ (remembering that $\lambda_1(\mu) = 1$ for any $N$), but we now can easily extend the argument to any excitation $j$ by total symmetry, given that the $B$'s all commute with one another. In other words, we can repeat the argument above for any $j$ by simply pulling out the relevant $B(\mu_{j})$ in \eqref{for-Bethe-eqs}. In all cases, no other term can interfere with the relevant one, and since they all are of the same type one arrives at (\ref{Bethe2}).

}
\end{appendix}

\end{document}